\documentclass[a4paper,11pt]{article}
\pdfoutput=1 

\usepackage{jheppub} 
\usepackage{amsmath}
\usepackage{pict2e}
\usepackage{physics}
\usepackage{amsfonts}
\usepackage{mathtools}
\usepackage{amssymb,bbm}
\usepackage{dsfont}
\usepackage{subcaption}
\usepackage{braket}
\usepackage{amsthm}
\usepackage{xcolor}
\usepackage{tikz}
\usepackage{ytableau}
\usetikzlibrary{hobby}
\usetikzlibrary{arrows.meta}
\usetikzlibrary{calc,arrows,cd,decorations.markings,snakes}
\tikzset{
->-/.style args={#1rotate#2}{decoration={markings, mark=at position #1 with {\arrow[scale=1.5,rotate = #2 ]{stealth}}}, postaction={decorate}}
}

\makeatletter
\newcommand{\farsquare}[2]{#1\,{\mathpalette\far@square{#2}}}
\newcommand{\far@square}[2]{%
  \mathop{\vcenter{\hbox{%
    \sbox\z@{$\m@th#1\sum$}%
    \setlength{\unitlength}{0.9\dimexpr\ht\z@+\dp\z@}%
    \begin{picture}(1,1)
    \roundjoin
    \polyline(0,0)(0,1)(1,1)(1,0)(0,0)(0,0.5)
    \end{picture}%
  }}}\limits_{#1#2}%
}
\makeatother

\newcommand{\nc}{\newcommand}
\nc{\rnc}{\renewcommand} 

\rnc{\a}{\alpha}
\rnc{\b}{\beta}
\nc{\g}{\gamma}
\rnc{\d}{\delta}
\nc{\e}{\epsilon}
\nc{\ee}{\varepsilon}
\nc{\z}{\zeta}
\nc{\f}{\phi}
\nc{\m}{\mu}
\nc{\n}{\nu}
\rnc{\r}{\rho}
\rnc{\k}{\kappa}
\rnc{\l}{\lambda}
\nc{\p}{\pi}
\nc{\s}{\sigma}
\rnc{\t}{\tau}
\nc{\w}{\omega}
\nc{\x}{\chi}
\nc{\F}{\Phi}
\rnc{\L}{\Lambda}

\nc{\1}{\mathbbm{1}}

\usepackage[draft,inline,nomargin,index]{fixme}
\FXRegisterAuthor{kr}{akr}{\color{blue}KR}

\title{\boldmath Non-invertible symmetries in $S_N$ orbifold CFTs and holography}

\author[a]{Michael Gutperle,}
\author[a]{Yan-Yan Li,}
\author[a]{Dikshant Rathore,}
\author[b]{and Konstantinos Roumpedakis}

\affiliation[a]{
Mani L. Bhaumik Institute for Theoretical Physics, Department of Physics and Astronomy,\\ University of California, Los Angeles, CA 90095, U.S.A.
}
\affiliation[b]{William H. Miller III Department of Physics and Astronomy, Johns Hopkins University,\\ 3400 North
Charles Street, Baltimore, MD 21218, U.S.A.
}

\emailAdd{gutperle@ucla.edu}
\emailAdd{yanyanli@ucla.edu}
\emailAdd{drathore@physics.ucla.edu}
\emailAdd{kroumpe1@jh.edu}

\abstract{We study non-invertible defects in two-dimensional $S_N$ orbifold CFTs. We construct universal defects which do not depend on the details of the seed CFT  and hence  exist in any orbifold CFT. Additionally, we investigate non-universal defects arising from the topological defects of the seed CFT. We argue that there exist universal defects that are non-trivial in the large-$N$ limit, making them relevant for the AdS$_3$/CFT$_2$ correspondence. We then focus on AdS$_3\times$S$^3\times \mathcal M_4$ with one unit of NS-NS flux and propose an explicit realization of these defects on the worldsheet.}

\begin{document} 
\maketitle
\flushbottom

\section{Introduction}
\label{sec:intro}

In recent years, generalized and non-invertible symmetries in various dimensions have been a very active field of investigation \cite{Chang:2018iay,Lin:2019hks,Thorngren:2019iar,Komargodski:2020mxz, Thorngren:2021yso, Nguyen:2021yld, Choi:2021kmx, Kaidi:2021xfk, Choi:2022zal, Cordova:2022ieu, Roumpedakis:2022aik,Bhardwaj:2022yxj,Damia:2022rxw,Arias-Tamargo:2022nlf,Hayashi:2022fkw,Kaidi:2022uux,Antinucci:2022eat, Choi:2022rfe, Choi:2022jqy,Bartsch:2022mpm, Heidenreich:2021xpr, Cordova:2022rer,Huang:2021zvu, Bashmakov:2022jtl,Benini:2022hzx, Damia:2022bcd,Lin:2022dhv, Niro:2022ctq,Cordova:2023ent, Rudelius:2020orz, Inamura:2021wuo, Nguyen:2021naa, Bhardwaj:2022lsg, Kaidi:2022cpf, Chen:2022cyw, Bashmakov:2022uek, Cordova:2022fhg, Choi:2022fgx,Yokokura:2022alv,Bhardwaj:2022kot,Bhardwaj:2022maz,Bartsch:2022ytj,Apte:2022xtu,Kaidi:2023maf,Koide:2023rqd,Damia:2023ses,Bhardwaj:2023ayw,vanBeest:2023dbu,Lawrie:2023tdz,Chen:2023czk,Cordova:2023bja,Antinucci:2023ezl,Benedetti:2023owa,Choi:2023pdp,Nagoya:2023zky,Okada:2024qmk,Kan:2024fuu} providing valuable insight into the dynamics of quantum field theories. At the core of these generalizations is the realization of symmetries as (extended) topological operators and vice versa. In this picture, a natural question is to find the holographic dual of these symmetry operators in the bulk. In theories with weakly coupled holographic duals, it has been proposed that boundary topological operators correspond to D-branes in the bulk  \cite{Apruzzi:2022rei,GarciaEtxebarria:2022vzq, Heckman:2022xgu, Heckman:2022muc,vanBeest:2022fss, Etheredge:2023ler, Bah:2023ymy,Dierigl:2023jdp, Bourget:2023wlb,Apruzzi:2023uma, Baume:2023kkf, Cvetic:2023pgm, Heckman:2024oot,Argurio:2024oym, Zhang:2024oas} which, in certain limits, can become topological.  
These proposals are based on anomaly inflow arguments. The goal of this paper is to put forward a proposal for the holographic dual of topological operators in a setup with a strongly coupled bulk theory using the AdS$_3$/CFT$_2$ correspondence. 

In two-dimensional conformal field theories (CFT), non-invertible symmetries have a long history starting with the work of Verlinde in rational CFTs \cite{Verlinde:1988sn}, the construction of topological defects in terms of projectors  \cite{Petkova:2000ip,Petkova:2001ag}, the construction of topological duality defects   \cite{Frohlich:2004ef,Frohlich:2006ch}, and the topological field theory formulation of RCFTs \cite{Fuchs:2002cm} to name a few. Orbifolds are a powerful way to obtain new CFTs from old ones \cite{Dixon:1985jw,Dixon:1986qv}.  In this paper, we focus on the so-called symmetric orbifold CFT, where one takes the tensor product of  $N$ copies of a seed CFT $\mathcal M$ and gauges the $S_N$ symmetric group \cite{Dijkgraaf:1996xw} that permutes the different copies. A remarkable feature of these theories is that they observe a large-$N$ factorization of correlation functions \cite{Jevicki:1998bm,Lunin:2000yv,Pakman:2009zz}, similar to the case of single-trace operators in large-$N$ gauge theories.  Proposals for the holographic dual of symmetric orbifold CFTs go back to \cite{Maldacena:1997re}, relating a type IIB $AdS_3\times S^3\times \mathcal{M}_4$ background, with $\mathcal{M}_4$ being either $K_3$ or $T^4,$ to the ${\cal N}=(4,4)$ supersymmetric orbifold CFT $\mathcal{M}_4^{\otimes N}/S_N$. This background allows for an exact worldsheet description with only NS-NS flux. However, it has been a challenge to identify the particular background which corresponds to the undeformed $\mathcal{M}_4^{\otimes N}/S_N$ orbifold \cite{Seiberg:1999xz}. Recently, such a dual was found for the  $AdS_3\times S^3\times \mathcal{M}_4$ background with the minimal $k=1$ unit of NS-fivebrane flux \cite{Gaberdiel:2017oqg,Gaberdiel:2018rqv,Eberhardt:2018ouy,Giribet:2018ada,Eberhardt:2019ywk}.

It is well known that gauging a group-like symmetry leads to a quantum symmetry in the gauged theory \cite{Vafa:1989ih, Tachikawa:2017gyf}. Hence, by construction, an orbifold CFT exhibits a universal set of symmetries arising by the gauging of a discrete group. In the case of interest, the gauging of the non-abelian $S_N$ group leads to a universal set of non-invertible symmetries. In this work, we look at these universal defects in two-dimensional symmetric orbifold CFTs. We construct them by applying the projector construction of Petkova and Zuber \cite{Petkova:2000ip}, and use the modular transformations of the partition function to identify the defect Hilbert spaces of operators on which line-defects can end. Additionally, we propose a realization of the defect operators in terms of non-gauge invariant operators under $S_N$. These defects, being universal, depend solely on the group theory structure of $S_N$ and not on the specifics of the seed CFT. To illustrate their properties, we examine the simplest example that exhibits non-invertible symmetries, i.e. $N=3$.

In symmetric orbifold CFTs, universal defects are labeled by the representations of $S_N$. We study these defects in the large-$N$ limit and analyze which representations remain non-trivial. We argue that these defects correspond to topological defects on the worldsheet and propose an explicit realization. While we verify some of their properties, further investigation is required for a complete realization.

The structure of this paper is as follows: In section \ref{sec2}, we review some background material, in particular non-invertible symmetries in RCFTs and the construction of 
symmetric orbifold CFTs. In section \ref{sec:universal}, we construct the universal topological line-defect operators in the symmetric orbifold CFTs that
realize the Rep$(S_N)$ fusion algebra. We use the modular transformation of the torus partition function with a defect inserted to determine the corresponding defect Hilbert spaces, and we construct the vaccum-state operators in terms of non-gauge invariant twist fields.
In section \ref{sec:s3example},  we present a detailed example of the $S_3$ symmetric orbifold. This is one of the simplest examples where non-invertible symmetries arise in CFTs.  We determine the lines and
defect operators, as well as the Ward identities for two and three-point correlators of twist operators.
In section \ref{sec:non-universal}, we discuss non-universal topological defects, starting with a nontrivial topological defect 
operator in the seed theory and constructing a ``maximally fractional'' defect, where all sectors are built on the same seed CFT defect 
in an $S_N$ invariant manner. We show that such defects are consistent, in the sense that they satisfy the Cardy-Petkova-Zuber \cite{Cardy:1989ir,Petkova:2000ip,Petkova:2001ag} conditions.   In section \ref{sec:large-N}, we 
discuss some aspects of the large-$N$ limit of these defects, particularly whether a non-invertible defect survives this limit. In section \ref{sec:non-inv-ads-cft}, we discuss the AdS/CFT dual of 
the symmetric orbifold   \cite{Eberhardt:2018ouy}, which states that the AdS dual is a type IIB string theory on 
$AdS_3\times S^3 \times \mathcal{M}_4$ with one unit of NS-NS fivebrane flux. We propose a worldsheet dual of the universal defects and discuss some of their properties. In section \ref{sec:Discussion}, we close with some discussion of our results and list some open questions and speculations. Some technical details of the $S_3$ example are relegated to appendix \ref{appendix:a}.

\section{Review}
\label{sec2}
In this section, we review well-known material which will play an important role in the main part of the paper. First, we discuss the construction of Verlinde lines in rational conformal field theories, introducing key concepts of non-invertible symmetries in two-dimensional CFTs. Second, we review the construction of the $\mathcal{M}^{\otimes N}/S_N$ symmetric orbifold CFT, for which we will construct the non-invertible symmetries in following sections.

\subsection{Non-invertible symmetries in rational CFTs}

\label{subsec:rcft}
Let us first look at the construction of topological line-defects in rational conformal field theories (RCFTs). We will restrict our attention to the case of diagonal RCFTs and the Verlinde lines \cite{Verlinde:1988sn} which are in one-to-one correspondence with the chiral primaries. The torus partition function of such  an RCFT is given by
\begin{align}
    Z(\tau,\bar{\tau}) = \sum_{i,j} n_{ij} \;  \chi_i(\tau) \bar{\chi}_j(\bar{\tau})~,
\end{align}
where $n_{ij} = \delta_{ij}$ and $\chi_i(\tau)$ denotes the character of a finite irreducible chiral algebra labeled by the representation $i.$ Under the modular $S$-transformation, the characters transform as $\chi_i(-1/\tau) = \sum_j S_{ij} \; \chi_j(\tau).$
The operator corresponding to a Verlinde line ${\cal I}_a$ associated to the chiral primary $\ket{\Phi_a}$ is given by 
\cite{Petkova:2000ip,Petkova:2001ag}
\begin{align}\label{rcft defect}
    {\cal I}_a = \sum_i \frac{S_{ai}}{S_{0i}} P_{i \bar{i}}~,
\end{align}
where $P_{i\bar{i}}$ is a projector acting on the space spanned by the $i$-th primary and its descendants:
\begin{align}
    P_{i\bar{i}} = \sum_{n \bar{n}} \ket{i,n} \otimes \ket{i, \bar{n}} \bra{i, n} \otimes \bra{i, \bar{n}}~.
\end{align}
Using the Verlinde formula \cite{Verlinde:1988sn},
\begin{align}
   n^{c}_{\;ab} =  \sum_i {S_{ai} \; S_{bi} \;  S_{ci}^* \over S_{0i}}~,
\end{align}
we can verify  that the fusion algebra of the Verlinde lines is isomorphic to that of the chiral primaries,
\begin{align}
    {\cal I}_a \; {\cal I}_b = \sum_c n_{\;ab}^c \; {\cal I}_c~, 
\end{align}
where $n_{\;ab}^c$ is the non-negative integer fusion coefficient,  which denotes the multiplicity with which ${\cal I}_c$ appears in the fusion product of the defects.
Given that the Verlinde lines act as projectors on the modules of the chiral algebra, which includes the Virasoro algebra, they commute with both the holomorphic and anti-holomorphic modes of the stress tensor,
\begin{align}
    L_{n} \; {\cal I} _a  &= {\cal I} _a \; L_n~, \quad  \quad \bar L_{n} \;  {\cal I} _a  = {\cal I} _a  \;\bar L_n~,
 \end{align}
and are therefore topological. Using the folding trick of \cite{Oshikawa:1996dj,Bachas:2001vj},  the topological defects have a realization as Ishibashi boundary states \cite{Ishibashi:1988kg,Cardy:1989ir} which characterize $\mathbb{Z}_2$ permutation branes 
\cite{Recknagel:2002qq,Drukker:2010jp,Cordova:2023qei}. If we start with the folded theory $\text{CFT}_{(1)} \cross \overline{\text{CFT}}_{(2)},$ the boundary state defining a $\mathbb{Z}_2$ permutation brane is given by
\begin{align}\label{doubled-a}
\mid B_a\rangle \rangle = \sum_ i {S_{ai}\over S_{0i} }   \sum_{n} \ket{i, n}_{(1)} \otimes \overline{\ket{i,n}}_{(2)}   \sum_{m}   \ket{i,m}_{(2)} \otimes \overline{\ket{i,m}}_{(1)}
~,
\end{align}
such that they obey the following Virasoro gluing conditions:
\begin{align}
\Big(L^{(1)} _k  -\bar L^{(2)} _{-k} \Big) \sum_n \ket{i, n}_{(1)} \otimes \overline{\ket{i,n}}_{(2)} &=0~, \nonumber \\
\Big(L^{(2)} _k  -\bar L^{(1)} _{-k} \Big) \sum_m \ket{i,m}_{(2)} \otimes \overline{\ket{i,m}}_{(1)} &=0~. \nonumber
\end{align}
The quantum dimension  of the topological defect operator is the eigenvalue of the defect acting on the CFT vacuum state,
\begin{align}
    {\cal I}_a \ket{0} = \langle {\cal I}_a \rangle \ket{0}  ~.
\end{align}
For the Verlinde lines we obtain $ \langle {\cal I}_a \rangle= S_{a0}/S_{00}$. It was pointed out in \cite{Chang:2018iay} that defects that have a quantum dimension not equal to one are in general associated with non-invertible symmetries. Note that in the boundary state representation (\ref{doubled-a}) the quantum dimension is obtained by $\langle {\cal I}_a \rangle=\langle 0\mid B_a\rangle \rangle $.

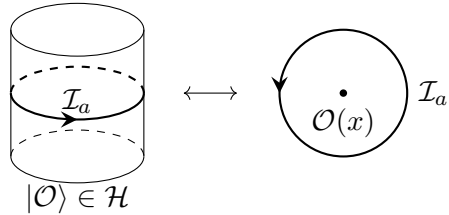
\begin{figure}[ht]
\centering
\begin{tikzpicture}[scale = 0.7]
\draw (0,0) ellipse (1.25 and 0.5);
\draw (-1.25,0) -- (-1.25,-2.4);
\draw (-1.25,-2.4) arc (180:360:1.25 and 0.5);
\draw [dashed] (-1.25,-2.4) arc (180:360:1.25 and -0.5);
\draw (1.25,-2.4) -- (1.25,0);  
\draw[thick, dashed] (1.25,-1.2) arc (360:180:1.25 and -0.5);
\draw[thick, ->- = .5 rotate 0] (-1.25,-1.2) arc (180:360:1.25 and 0.5);
\node at (0,-1.35) {${\cal I}_a$};
\node at (0,-3.2) {$ \ket{{\cal O}} \in {\cal H} $};
\draw[<->] (2,-1.2) -- (3,-1.2);
\draw[thick, ->- = .5 rotate 0] (5,-1.2) circle (1.2cm);
\node at (6.7,-1.2) {${\cal I}_a$};
\fill (5,-1.2) circle (2pt) node[below=2pt] {${\cal O}(x)$};
\end{tikzpicture}
\caption{Using the state-operator correspondence to depict the action of the line-defect on an operator in the untwisted Hilbert space.}
\label{fig:verlindeaction}
\end{figure}

Operators like \eqref{rcft defect} satisfying the Cardy-Petkova-Zuber \cite{Cardy:1989ir,Petkova:2000ip,Petkova:2001ag} conditions can be viewed as topological line-defects. The action of these line-defects can be visualized on the complex plane mapped from the cylinder as in Figure \ref{fig:verlindeaction}, where we insert the defect at a particular point in time. We refer to the Hilbert space defined on $S^1$ for every point in time as the \textit{untwisted} Hilbert space.

The corresponding torus partition function is given by inserting the defect inside the trace over the Hilbert space,
\begin{align}
    Z^{a}(\tau, \bar{\tau}) = \tr_{{\cal H}} [{\cal I}_a \; q^{L_0 - \tfrac{c}{24}} \; \bar{q}^{\bar{L}_0 - \tfrac{c}{24}}]~,
\end{align}
where $q = e^{2 \pi i \tau}$ with $\tau$ being the torus modular parameter. After performing a modular $S$-transformation, the line-defect now extends in the time direction, and at every point in time we obtain the \textit{defect Hilbert space}, ${\cal H}_{{\cal I}_a}.$ The torus partition function is now modified to trace over this defect Hilbert space corresponding to the defect ${\cal I}_a,$
\begin{align}
    Z_{a}(-\tfrac{1}{\tau}, -\tfrac{1}{\bar{\tau}}) = \tr_{{\cal H}_{{\cal I}_a}} [ \tilde{q}^{L_0 - \tfrac{c}{24}} \; \bar{\tilde{q}}^{\bar{L}_0 - \tfrac{c}{24}}]~,
\end{align}
where $\tilde{q} = e^{-2 \pi i /\tau}.$ Using the $S$-transformation of the characters, and the fact that the defect commutes with the chiral vertex algebra, we can expand the defect Hilbert space in terms of the characters. The defect partition function can thus be written as
\begin{align}
    Z_{a}(\tau,\bar{\tau}) = \sum_{b,c} N_{ab}^c \;  \chi_b(\tau) \bar{\chi}_c(\bar{\tau})~,
\end{align}
where the non-negative integers $N_{ab}^c$ encode the degeneracy of the states in ${\cal H}_{{\cal I}_a}.$ Note that even though we started with a diagonal theory, the defect partition function is not necessarily diagonal, i.e. $N_{ab}^c$ need not be a diagonal matrix. This tells us that states in ${\cal H}_{{\cal I}_a}$ have non-integer spins and correspond to non-local operators attached to the corresponding line ${\cal I}_a.$

\begin{figure}[ht]
\centering
\begin{tikzpicture}[scale = 0.7]
\draw (0,0) ellipse (1.25 and 0.5);
\draw (-1.25,0) -- (-1.25,-2.4);
\draw (-1.25,-2.4) arc (180:360:1.25 and 0.5);
\draw [dashed] (-1.25,-2.4) arc (180:360:1.25 and -0.5);
\draw (1.25,-2.4) -- (1.25,0);  
\draw[thick, ->- = .5 rotate 0]  (0,-2.9) -- (0,-0.5);
\node at (0.5,-1.2) {${\cal I}_a$};
\node at (0,-3.2) {$ \ket{\Psi} \in {\cal H}_{{\cal I}_a} $};
\draw[<->] (2,-1.2) -- (3,-1.2);
\draw[thick, ->- = .5 rotate 0] (4,-2.7) -- (4,-0.5) ;
\node at (4.5,-1.2) {${\cal I}_a$};
\fill (4,-2.7) circle (2pt) node[below=2pt] {$\Psi(x)$};
\end{tikzpicture}
\caption{States in the defect Hilbert space ${\cal H}_{{\cal I}_a}$ correspond to non-local operators attached to the line-defect ${\cal I}_a.$}
\label{fig:defectline}
\end{figure}
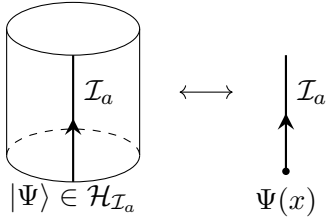

An important feature of non-invertible line-defects is the ``lasso'' effect \cite{Chang:2018iay} where, when a topological line-defect is moved past a local operator, one obtains in addition to the symmetry action on the local operator another topological line ending on a defect operator and a junction of topological line operators (see Figure \ref{fig:non.inv.lines}).
The mathematical structure describing symmetry line-defects,  defect operators, and their manipulations is called a tube algebra \cite{Lin:2022dhv}.

\begin{figure}[ht]
  \[
    \begin{tikzpicture}[scale = 1, baseline = 0]
        \coordinate (o) at (0,0);
        \coordinate (d) at (0.5,-1);
        \coordinate (m) at (0.5,0);
        \coordinate (u) at (0.5,1);
        \draw[fill] (o) circle [radius = .05];
        \node[anchor = south] at (o) {$\mathcal{O}$};
        \draw[thick, ->- = .55 rotate 0] (d) -- (u);
        \node[anchor = west] at (m) {$\;{\cal I}$};
    \end{tikzpicture} = \frac{1}{\mathrm{dim}\,{\cal I}} \left(
    \begin{tikzpicture}[scale = 1, baseline = 0]
        \coordinate (o) at (0.7,0);
        \coordinate (d) at (0,-1);
        \coordinate (m) at (0,0);
        \coordinate (u) at (0,1);
        \draw[fill] (o) circle [radius = .05];
        \node[anchor = south] at (o) {${{\cal I} \cdot \mathcal{O}}$};
        \draw[thick, ->- = .55 rotate 0] (d) -- (u);
        \node[anchor = east] at (m) {${\cal I}\;$};
    \end{tikzpicture} \right) +
    \begin{tikzpicture}[scale = 1, baseline = 0]
        \coordinate (o) at (1,0);
        \coordinate (d) at (0,-1);
        \coordinate (m) at (0,0);
        \coordinate (u) at (0,1);
        \draw[fill] (o) circle [radius = .05];
        \node[anchor = south] at (o) {$\mathcal{O}_{\mathrm{defect}}$};
        \draw[thick, ->- = .55 rotate 0] (d) -- (m);
        \draw[thick, ->- = .55 rotate 0] (m) -- (u);
        \draw[->- = .5 rotate 0,dashed] (m) -- (o);
        \node[anchor = east] at ($(m)!.5!(d)$) {${\cal I}$};
        \node[anchor = east] at ($(m)!.5!(u)$) {${\cal I}$};
    \end{tikzpicture}
\]
\caption{Lasso action of a topological line-defect.}
\label{fig:non.inv.lines}
\end{figure}

\subsection{Symmetric Orbifold CFT}
\label{sec:symorb}

In this section, we review the symmetric orbifold CFT. Consider $N$ copies of a CFT $\mathcal{M}$ (not necessarily rational). The tensor product CFT has an $S_N$ symmetry that permutes the $N$ copies of the theory. We construct the $S_N$ orbifold CFT by gauging this permutation symmetry. Let's denote the primary operators of the tensor product theory by ${\cal O}_I$ where $I= 1,2, \dots, N$. The spectrum of the orbifold theory consists of the untwisted sector constructed by taking gauge invariant combinations of operators in the tensor product theory such as
\begin{equation}
    \sum_{I=1}^N {\cal O}_I, \quad  \sum_{I,J=1}^N {\cal O}_I {\cal O}_J,  \dots ~.
\end{equation}
Modular invariance additionally requires twisted sectors. We introduce the twist operators $\s_g$ labeled by the elements of $S_N$ by
\begin{equation}
    {\cal O}_I(e^{2\pi i }z) \; \s_g(0) = {\cal O}_{g\cdot I}(z)\; \s_g(0)~. \label{eq:twist_def}
\end{equation}
These operators introduce non-trivial holonomies for the operators of the seed theory. However, the operators $\s_g$ are not gauge invariant. To define gauge invariant operators we define  \cite{Dijkgraaf:1989hb, Lunin:2000yv, Pakman:2009zz}

\begin{equation}\label{eq:g-inv_def}
    \s_{[g]} \propto \sum_{h \in S_N}  \s_{h^{-1} g h}~,
\end{equation}
labeled by the conjugacy classes $[g]$ of $S_N$. The proportionality constant can be fixed by normalizing the two-point function of these operators \cite{Pakman:2009zz}. A conjugacy class $[g]$  is fixed by giving the number $l_i$ of cycles of length $i=1,2,\cdots N$.
\begin{align}\label{cyclemg}
    [g] = { 1} ^{l_1} \cdot  { 2}^{l_2} \cdots {N}^{l_N}~,\quad \sum_k k \; l_k=N~.
\end{align}
Excited states are created by acting with operators of the seed theory on twisted states. For example, consider the non-gauge invariant twist operator $\s_g$ where $g$ is a cycle of length $n$. Given \eqref{eq:twist_def}, the operators of the seed theory have fractional modes

\begin{equation}
    {\cal O}_I(z) = \frac{1}{n} \sum_m ({\cal O}_I)_{\frac{m}{n}} z^{-\frac{m}{n}-h_{{\cal O}}} e^{-2 \pi i \frac{m}{n}(I-1)}~.
\end{equation}
The modes with fractional indices correspond to excited twist operators.  Hence, in the twisted sector corresponding to an element of $S_N$ with a single cycle of length $n$, there are $n$ subsectors (the vacuum twist operator $\s_n$ and exited states dressed with fractional modes of the operators of the seed theory). More generally, in a twisted sector created by $\s_{g}$, twist operators are labeled by the centralizer of $g$. Thus, the torus  partition function can be written as follows \cite{Dijkgraaf:1989hb,Klemm:1990df}
\begin{align}
Z &= {1\over |G|} \sum_{gh=hg}   \farsquare{h}{g} \; (\tau,\bar \tau)~,
\label{toruspf}
\end{align}
where $g$ denotes the twisted sector and $h$ an element of its centralizer. Each term in the sum is equal to  
\begin{align}
 \farsquare{h}{g} \; (\tau,\bar \tau) = {\rm tr}_{{\cal H}_g} \Big( h \; e^{2\pi i \tau (L_0-{c\over 24}) } \; e^{-2\pi i \bar \tau (\bar L_0-{c\over 24}) } \Big)~.
\end{align}
The summation of elements $h\in S_N$ that commute with $g$ imposes the projection onto states that are invariant under the centralizer $C[g]$. Here   $|G|=N!$ is the order of $S_N$.
Using the fact that the torus partition function in the individual sectors transforms as follows under modular transformations $\tau\to -{1\over \tau}$ 
 \begin{align}\label{modular-a1}
  \farsquare{h}{g} (\tau, \bar \tau) \to  \farsquare{g^{-1}}{h} (-{1\over\tau},-{1\over  \bar \tau}) ~,
 \end{align}
it is straightforward to check that the orbifold partition function is modular invariant. 

The spectrum of the twisted sectors can be obtained by expanding the partition function. For a twisted sector labelled by a cycle $[g]$ in (\ref{cyclemg}), the conformal dimensions $h_{CFT}$ of the universal twisted sector ground state is given by
\begin{align}\label{twist-dim}
    h_{CFT}= \sum_{k=1}^N  l_k  {c_{\rm seed}\over 24}\big(k-{1\over k}\big)~.
\end{align}
These operators do not depend on the details of the seed CFT.
In addition, there are twisted sector states built on primaries in the seed theory with $h_{\rm seed}>0$ as well as (fractional) Virasoro descendants, but these states will not play an important role in the following.


Another way \cite{Dijkgraaf:1996xw,Maldacena:1999bp} to look at the partition function is via a grand canonical
 generating function. The modular invariant partition function of the seed theory  is  denoted as $Z(\tau,\bar \tau)$
 \begin{align}
 \sum_N p^N Z_{N} = \exp [ \sum_{k=1}^\infty p^k T_k Z(\tau,\bar \tau)]~,
 \label{eq:Grand_can}
 \end{align} 
where $T_k$ is the Hecke operator
\begin{align}
T_k Z(\tau,\bar \tau) = {1\over k} \sum_{i|k} \sum_{j=0}^{i-1} Z\Big( {k \tau\over i^2}+ {j\over i}, {k \bar \tau\over i^2}+ {j\over i}\Big)~.
\end{align}
Correlation function of twist operators can be calculated by going to the covering space \cite{Dixon:1986qv, Lunin:2000yv, Pakman:2009zz, Roumpedakis:2018tdb, Dei:2019iym}. Suppose we want to calculate a correlation function of twist operators. Due to \eqref{eq:twist_def}, correlation functions of non-gauge invariant operators have branch cuts. The union of the different sheets forms the covering space. Although we consider the orbifold theory on $S^2$, the covering space does not necessarily have zero genus; its genus is determined by the Riemann–Hurwitz formula \cite{Lunin:2000yv}.

\section{Universal defects in symmetric orbifolds}
\label{sec:universal}
We construct the orbifold CFT starting with a tensor product of $N$ copies of a seed theory $\mathcal{M}$ and gauge the $S_N$ symmetry that permutes the $N$ copies. The gauged theory exhibits a quantum symmetry Rep$(S_N)$\footnote{Intuitively, this symmetry is generated by the Wilson lines of the $S_N$ gauge field.} \cite{Vafa:1989ih, Tachikawa:2017gyf, Thorngren:2021yso}. In this section, we construct these universal topological defects using the projector construction of Petkova and Zuber \cite{Petkova:2000ip}. In particular, the details of the seed CFT will not play a role in the sense that the defect can be thought of as being built on the trivial topological defect in the seed CFT, which always exists. In addition to these defects, there may be other topological defects coming from the defects of the seed theory itself. We discuss non-universal defects, which are based on nontrivial topological defects in the seed theory, in section \ref{sec:non-universal}.

\subsection{Construction}

Utilizing the folded permutation brane construction for a topological defect, a universal defect corresponds to the following boundary state in the folded $\mathcal M^{\otimes N}/S_N \times \mathcal M^{\otimes N}/S_N$ CFT
 \begin{align}\label{boundary-state}
 \mid B_R\rangle \rangle &= \sum_{[g]}  \chi_R([g])  \sum_n  \mid n\rangle_{(1),[g]} \otimes \overline{ \mid n\rangle}_{(2),[g]} \sum_m  \mid m\rangle_{(2),[g]}  \otimes \overline{ \mid m\rangle}_{(1),[g]} ~,
 \end{align}
 Here $[g]$ denotes the twisted sector in the $\mathcal M^{\otimes N}/S_N$ orbifold labeled by the conjugacy class of $S_N$. The summation over $n,m$ denotes the sum over all states in the $[g]$-twisted sector including the sum over primaries of the seed theory as well as  Virasoro descendants producing an Ishibashi boundary state \cite{Ishibashi:1988kg}. The folded boundary state, and thus the defect, is labeled by the choice of representation $R$ of S$_N$ with $\chi_R([g])$ being the character of the conjugacy class $[g]$ in the representation $R$.  Consequently, the boundary state satisfies
 \begin{align}\label{boundary-a}
     \Big(L^{(1)} _n-\bar L^{(2)}_{-n}\Big) \mid B_R\rangle \rangle &=  \Big(L^{(2)} _n-\bar L^{(1)}_{-n}\Big) \mid B_R\rangle \rangle =0~,
 \end{align}
where $L^{(i)}_n, i=1,2$ are the modes of the stress tensor of the two copies of the $\mathcal M^{\otimes N}/S_N$ orbifold after folding. After unfolding, the two copies of the CFT living on half-spaces are separated by the defect operator 

 \begin{align}\label{td-projector}
{\cal I} _R&=  \sum_{[g]}  \chi_R([g]) {P}_{[g]}\bar { P}_{[g]} ~.
\end{align} 
 The defect is expressed in terms of the identity projectors in the $[g]$-twisted sector satisfying ${ P}_{[g]}^2={P}_{[g]}$ and $ { P}_{[g]}{ P}_{[g']}=0$ if $[g]\neq [g']$. It is clear from the property of the projectors or the unfolding of the boundary state condition (\ref{boundary-a}) that the defect is topological, i.e. it commutes with the holomorphic and anti-holomorphic modes of the stress tensor separately.
 \begin{align}\label{td-state}
     L_{n} \;{\cal I} _R  &= {\cal I} _R \; L_n~, \quad  \quad \bar L_{n} \; {\cal I} _R  = {\cal I} _R \;  \bar L_n~.
 \end{align}
Using the  properties of the projection operators, one can calculate the fusion product of two defect operators labeled by two representations $R_{1,2}$

 \begin{align}
 {\cal I}_{R_1} {\cal I}_{R_2} 
 &= \sum_{[g]} \chi_{R_1} ([g]) \chi_{R_2} ([g]) {P}_{[g]}\bar { P}_{[g]}\nonumber\\
 &=  \sum_{ R_3} {1\over |G|} \sum_g \chi_{R_1} (g) \chi_{R_2}(g)\chi_{R_3}(g)^*  {\cal I}_{R_3} \nonumber \\
 &=\sum_{ R_3}  N_{R_1 R_2}^{R_3}  {\cal I}_{R_3}~.
  \end{align}
Here $N_{R_1 R_2}^{R_3}$ are the Kronecker coefficients of the decomposition of the tensor product $R_1\otimes R_2$ into representations $R_3$.  Since $S_N$ for $N>2$ is a non-abelian group, the topological line-defects include non-invertible ones. This can also be seen from the fact that acting on a state in the $[g]$-twisted sector one has
\begin{align}\label{twist-act}
    {\cal I}_R \mid \Phi_{[g]} \rangle=\chi_R([g]) \mid \Phi_{[g]} \rangle~.
\end{align}
In particular, this implies that the quantum dimension of a defect $ {\cal I}_R$ is given by
$\langle {\cal I}_R \rangle = \chi_R( 1)=dim_R$, i.e. the dimension of the representation. In general, for $S_N$ with  $N>2,$ apart from the trivial and the alternating (or sign) representation, all other representations have dimensions greater than one, hence giving rise to non-invertible defects.

\subsection{Modular invariance and defect operators}\label{defectH}

In this section, we use modular invariance to identify the operators in the defect Hilbert space which can end on topological lines. The starting point is to insert a defect ${\cal I}_R$ along the spatial cycle in the torus partition function

\begin{align}
    Z^R (\tau,\bar \tau)&= {1\over |G|} \sum_{gh=hg} tr_{{\cal H}_g} \Big( h \; e^{2\pi i \tau (L_0-{c\over 24}) } \; e^{-2\pi i \bar \tau (\bar L_0-{c\over 24}) }  \; {\cal I}_R\Big) \nonumber\\
    &= {1\over |G|} \sum_{gh=hg}  \chi_R([g])\;  \farsquare{h}{g} \; (\tau, \bar \tau)~.
    \label{twsitedpf}
\end{align}
The defect Hilbert space can be obtained by performing a modular transformation $\tau \to -{1\over \tau}=\tau',$ which exchanges the spatial and temporal cycles.   Using the modular transformation (\ref{modular-a1}), the defect partition function is given by
\begin{align}
Z_R (\tau',\bar \tau')  = {1\over |G|} \sum_{gh=hg}  \chi_R([g])\;  \farsquare{g}{h} \; (\tau',\bar \tau')~,
\end{align}
where we used the fact that the characters of $S_N$ are real.

In the orbifold partition function, we decompose the elements $h,g$ which commute $gh=hg$ by labeling the element $g$ by its conjugacy class and $h$ by the element of the centralizer. However, one can do this the other way around, where $h$ labels conjugacy classes and $g$ labels the centralizer of $h$.   

The simplest and most relevant case to the large-$N$ limit is when  $h$ corresponds to a single cycle of length $[w]$. The centralizer of such an element is $\mathbbm{Z}_w\times S_{N-w}$.  The group $\mathbbm{Z}_w$ contains elements $1, \omega, \omega^2, \cdots \omega^{w-1}$, where $\omega$ is the $w$-th root of unity. The defect partition function corresponding to the single $w$-twisted cycle can be written as
\begin{align}
Z_{R,{[w]}} = {1\over \dim C[w]} \; \chi_R( C[w])  \; \farsquare{C[w]}{[w]}~.
\end{align} 
Using the decomposition of $C[w]=\mathbbm{Z}_w \times S_{N-w}$ and $ \dim C[w]= w (N-w)!,$ one gets
\begin{align}\label{zrw}
Z_{R,{[w]}} = {1\over (N-w)!}\sum_{g\in S_{N-w}} {1\over w} \Big( \chi_R  ( [1_w \times g])\;   \farsquare{1_w\times g }{[w]}+ \sum_{k=1}^{w-1} \chi_R  ( [\omega^k \times g])\;  \farsquare{\omega^k \times g }{[w]}\Big)~.
\end{align}



To determine the number of defect operators, we want to pick out the ground state contribution of the $w$-twisted sector from \eqref{zrw},  which has the conformal dimension
\begin{align}\label{tgstate}
    h=\bar h ={c_{\rm seed} \over 24}\big( w -{1\over w}\big)~.
\end{align}
As far as the  $S_{N-w}$ permutation is concerned, the sector $[w]$ is untwisted. Consequently, the insertion of $g\in S_{N-w}$ contributes the same factor to the ground state in the defect partition function \eqref{zrw}.
The same holds for the $1_w$ and $\omega^k, k=1,2,\cdots w-1$ insertions since these enforce the $h-\bar h\in \mathbb{Z}$ projection, which is automatically satisfied for the twisted ground state \eqref{tgstate}.
This implies that 
the number of $w$-twisted defect ground states is simply obtained by setting $\Box =1$ for all terms  in  the partition function \eqref{zrw}, giving
\begin{align}\label{n-twista}
n^R_{w}= {1\over {(N-w)!} } \sum_{h'\in S_{N-w}} {1\over w}  \sum_{k=0}^{w-1} \chi_R([ \omega^k \times  h']) ~,
\end{align}
where we included  $\omega^0=1_w$.  A  vanishing  $n^R_{w}$ implies that the ${\cal I}_R$ defect cannot end on a bare twist operator in the $w$-twisted sector. 


One way to understand the defect operators at the endpoints of topological line-defects is to view these line-defects as open Wilson lines ending on non-gauge invariant operators \cite{Tachikawa:2017gyf}. 
A natural proposal for such operators involves using characters to project onto the representation $R$
\begin{align}\label{non-gauge-inv}
    \s^R_{g} \equiv N_R \sum_{h \in S_N} \chi_R (h) \; \s_{h^{-1} g h}~
,
\end{align}
where we pick an element $g$ in the conjugacy class that labels the twisted sector and we focus again on a single cycle of length $w$.   We propose that the ${\cal I}_R$ line-defect can end on a defect operator in the $[g]$-twisted sector if (\ref{non-gauge-inv}) does not vanish. We can decompose the sum over $h\in S_N$ in (\ref{non-gauge-inv}) into a sum over the centralizer of $g$ and the rest
\begin{align}\label{non-gauge-inv2}
    \s^R_{g} &\equiv N_R \Big( \sum_{h'\in S_{N-w}}    \sum_{k=0}^{w-1} \chi_R([ \omega^k \times  h']) \s_g+ \cdots \Big)~,
\end{align}
where the dots denote summation over other group elements in the conjugacy class of $g$. 
Note that the coefficient multiplying $\s_g$ in (\ref{non-gauge-inv2}) is proportional to (\ref{n-twista}).  Consequently, the vanishing of one implies the vanishing of the other and the criterion for the existence of a defect operator ending on a line-defect is the same.  The vanishing of all the terms on the right-hand side of (\ref{non-gauge-inv}) can be established by conjugating the group element $g$ and using the fact that the characters take the same value on all elements of the same conjugacy class. 

It should be noted that for $S_N$ with $N>4$, the degeneracy of the twisted sector ground states (\ref{n-twista}) can be larger than one.  For the non-gauge invariant operators in  (\ref{non-gauge-inv}), this can be understood by the fact that the action of $S_N$ on all elements $g$ in a fixed conjugacy class by conjugation $g\to h^{-1} g h$ determines a (generally) reducible representation. The value $n_w^R$ gives the number of times the irreducible representation $R$ appears in this representation.

In the next section we will further test this construction by evaluating correlation functions of defect operators and check the Ward identities following from the non-invertible symmetry Rep$(S_3)$.

\section{Example: $S_3$ symmetric orbifold}
\label{sec:s3example}

We now consider a simple example, the $S_3$ orbifold theory of a seed CFT $\mathcal{M}$ with central charge $c$. Upon gauging the $S_3$ symmetry, a Rep$(S_3)$ quantum symmetry emerges. The objects of Rep$(S_3)$ symmetry are topological lines which are labeled by the irreducible representations of $S_3$.

The symmetric group $S_3$ is a non-abelian group consisting of permutations of three elements. The group-elements of $S_3$ are
\begin{align}
    S_3 = \{ e, (12), (23), (31), (123), (132) \}~,
\end{align}
where $e$ is the identity permutation, $(12)$ is a 2-cycle representing the permutation $1 \rightarrow 2 \rightarrow 1,$ $(123)$ is a 3-cycle representing the permutation $1 \rightarrow 2 \rightarrow 3 \rightarrow 1$ and so on. There are three conjugacy classes
\begin{align}
    [e] = \{e\}, \; \; \; [2] = \{(12), (13), (23)\}, \; \; \; [3] = \{(123), (132)\}~.
\end{align}
The character table of $S_3$ is given below
\begin{align}
\begin{array}{l|rrr}
& [e]& [2] & [3]\\ \hline
\chi_e & 1 & 1 & 1\\
\chi_{A} & 1 & -1 & 1\\
\chi_S & 2 & 0 & -1\\
\end{array}
\label{characterS3}
\end{align}
where $e$ denotes the trivial representation, $A$ denotes the alternating representation, and $S$ denotes the standard representation.

In this case, the fusion rules of the universal defects defined in \eqref{td-projector} follow from the above character table and read
\begin{align}
    {\cal I}_A \cross {\cal I}_A &= 1, \nonumber \\ 
    {\cal I}_S \cross {\cal I}_A &= {\cal I}_A \cross {\cal I}_S = {\cal I}_S, \nonumber \\
    {\cal I}_S \cross {\cal I}_S &= 1 + {\cal I}_A + {\cal I}_S~. 
\end{align}
We see that the line-defect associated with the standard representation is non-invertible. 
Using \eqref{eq:g-inv_def}, the gauge invariant ground state operators are
\begin{align}\label{eq:g-inv_def_S3}
    \s_{[2]} &= \frac{1}{\sqrt{3}} \Big( \s_{(12)} + \s_{(23)} + \s_{(13)} \Big)
    ~, \nonumber \\
    \s_{[3]} &= \frac{1}{\sqrt{2}} \Big( \s_{(123)} + \s_{(132)} \Big)~,
\end{align}
where the proportionality constants were fixed by requiring that the two-point functions be normalized to $1.$ The action of the defects on these operators is summarized in Figure \ref{fig:td-s3}.

\begin{figure}[ht]
\centering
\begin{align}
&
\begin{tikzpicture}[scale=0.7]
\draw[thick, red,dashed] (0,0) circle (1.2cm);
\fill (0,0) circle (2pt) node[below=2pt] {$\sigma_{[2]}$};
\node at (2,0) {$=$};
\node at (3.5,0) {$- \;  \sigma_{[2]}$};
\end{tikzpicture} \; \; \;  \; \; \; \; 
\begin{tikzpicture}[scale=0.7]
\draw[thick, red, dashed] (0,0) circle (1.2cm);
\fill (0,0) circle (2pt) node[below=2pt] {$\sigma_{[3]}$};
\node at (2,0) {$=$};
\node at (3,0) {$\sigma_{[3]}$};
\end{tikzpicture} \nonumber \\ \nonumber \\
&
\begin{tikzpicture}[scale=0.7]
\draw[thick] (0,0) circle (1.2cm);
\fill (0,0) circle (2pt) node[below=2pt] {$\sigma_{[2]}$};
\node at (2,0) {$=$};
\node at (3,0) {$0$};
\end{tikzpicture} \; \; \; \; \; \; \; \; \; \; \; \; \; \;
\begin{tikzpicture}[scale=0.7]
\draw[thick] (0,0) circle (1.2cm);
\fill (0,0) circle (2pt) node[below=2pt] {$\sigma_{[3]}$};
\node at (2,0) {$=$};
\node at (3.5,0) {$ - \; \sigma_{[3]}$};
\end{tikzpicture} 
\nonumber
\end{align}
\caption{Action of the universal line-defects. The red line denotes the invertible sign defect $\mathcal I_A$ while the black line denotes the standard defect $\mathcal I_S$.}
\label{fig:td-s3}
\end{figure}

Next, let us discuss the defect Hilbert spaces. From \eqref{non-gauge-inv}, we observe that there are no two-cycle ground states in ${\cal H}_{{\cal I}_A}$ and no three-cycle ground states in ${\cal H}_{{\cal I}_S}.$ In the appendix, we verify this counting by analyzing the partition function. The twisted ground states in the defect Hilbert space are 
\begin{align}\label{eq:non-gauge_def_S3}
    \s^S_{[2]} &= N_S \; \Big(2 \s_{(12)} - \s_{(23)} -\s_{(13)} \Big)~, \nonumber \\
    \s^A_{[3]} &= N_A \; \Big(\s_{(123)} - \s_{(132)} \Big)~.
\end{align}
Here, without loss of generality, we choose $\sigma_{(12)}$ and $\sigma_{(123)}$ as representatives of $[2]$ and $[3]$ respectively. 
The constants $N_A, N_S$ can be fixed by normalizing the two-point correlators of the non-gauge invariant operators with the line-defects inserted to 1. For example, demanding that
\begin{align}
\begin{tikzpicture}
\node at (0,0) {$ \langle \; \sigma^S_{[2]} \hspace{1.5cm}  \sigma^S_{[2]} \; \rangle = N_S^2 \; \Big( 4 \langle \s_{(12)} \s_{(12)} \rangle + \langle \s_{(23)} \s_{(23)} \rangle  + \langle \s_{(13)} \s_{(13)} \rangle   \Big) = 6 N_S^2 = 1~, $};
\fill (-5.75,0) circle (2pt);
\fill (-4.6,0) circle (2pt);
\draw[thick] (-5.75,0) -- (-4.6,0);
\end{tikzpicture}
\end{align}
gives us $N_S^2 = 1/6.$ Similarly, we get $N_A^2 = - 1/2.$ To show this, we used that the two-point correlator vanishes unless the permutation corresponding to the second twist operator is the inverse of the first \cite{Lunin:2000yv}. In the above correlator, we implicitly assumed that insertions of the two operators are at the poles of the sphere.

Next, we determine the coefficients in the lasso diagrams. Since the defects are topological, the scaling dimension does not change and the two possible diagrams are shown in Figure \ref{fig:lasso}. In the following, we fix the coefficients $\alpha$ and $\beta$.  

\begin{figure}[ht]
\begin{align}
&
\begin{tikzpicture}[scale=0.7]
    \draw[thick] (0,0) circle (1.2cm);
    \fill (0,0) circle (2pt) node[above] {$\sigma_{[2]}$};
    \draw[thick] (1.2,0) -- (3.2,0);
    \node at (5,0) {$=$};
    \node at (6.8,0) {$\alpha$};
    \fill (7.6,0) circle (2pt) node[above] {$\sigma^S_{[2]}$};
    \draw[thick] (7.6,0) -- (9.6,0);
\end{tikzpicture}
\nonumber \\ \nonumber \\
&
\begin{tikzpicture}[scale=0.7]
    \draw[thick] (0,0) circle (1.2cm);
    \fill (0,0) circle (2pt) node[above] {$\sigma_{[3]}$};
    \draw[thick, red,dashed] (1.25,0) -- (3.2,0);
    \node at (5,0) {$=$};
    \node at (6.8,0) {$\beta$};
    \fill (7.6,0) circle (2pt) node[above] {$\sigma^A_{[3]}$};
    \draw[thick, red,dashed] (7.65,0) -- (9.6,0);
\end{tikzpicture} \nonumber
\end{align}
\caption{Lasso diagrams where the right-hand side is obtained after shrinking the standard defect loop which produces a defect operator.}
\label{fig:lasso}
\end{figure}

\begin{figure}[ht]
\begin{align}
&
\begin{tikzpicture}[scale=0.6]
    \draw[thick] (0,0) .. controls (0.75,1) .. (0,2);
    \draw[thick] (2,0) .. controls (1.25,1) .. (2,2);
    \node at (3,1) {\Large $=$};
    \node at (4,1) {\LARGE $\frac{1}{2}$};
    \draw[thick] (5,0) .. controls (6,0.75) .. (7,0);
    \draw[thick] (5,2) .. controls (6,1.25) .. (7,2);
    \node at (7.8,1) {\Large $+$ \LARGE $\frac{1}{2}$};
    \draw[thick] (8.8,0) .. controls (9.8,0.75) .. (10.8,0);
    \draw[thick] (8.8,2) .. controls (9.8,1.25) .. (10.8,2);
    \draw[thick,red,dashed] (9.8,1.45) -- (9.8, 0.55);
    \node at (11.8,1) {\Large $+$ \LARGE $\frac{1}{\sqrt{2}}$};
    \draw[thick] (12.8,0) .. controls (13.8,0.75) .. (14.8,0);
    \draw[thick] (12.8,2) .. controls (13.8,1.25) .. (14.8,2);
    \draw[thick] (13.8,1.45) -- (13.8, 0.55);
\end{tikzpicture}
\nonumber \\ \nonumber \\
&
\begin{tikzpicture}[scale=0.6]
    \draw[thick] (0,0) .. controls (0.75,1) .. (0,2);
    \draw[thick] (2,0) .. controls (1.25,1) .. (2,2);
    \draw[thick,red,dashed] (0.55,1) -- (1.45,1);
    \node at (3,1) {\Large $=$};
    \node at (4,1) {\LARGE $\frac{1}{2}$};
    \draw[thick] (5,0) .. controls (6,0.75) .. (7,0);
    \draw[thick] (5,2) .. controls (6,1.25) .. (7,2);
    \node at (8,1) {\Large $+$ \LARGE $\frac{1}{2}$};
    \draw[thick] (9,0) .. controls (10,0.75) .. (11,0);
    \draw[thick] (9,2) .. controls (10,1.25) .. (11,2);
    \draw[thick,red,dashed] (10,1.45) -- (10, 0.55);
    \node at (11.8,1) {\Large $-$ \LARGE $\frac{1}{\sqrt{2}}$};
    \draw[thick] (12.8,0) .. controls (13.8,0.75) .. (14.8,0);
    \draw[thick] (12.8,2) .. controls (13.8,1.25) .. (14.8,2);
    \draw[thick] (13.8,1.45) -- (13.8, 0.55);
\end{tikzpicture}
\nonumber \\ \nonumber \\
&
\begin{tikzpicture}[scale=0.6]
    \draw[thick] (0,0) .. controls (0.75,1) .. (0,2);
    \draw[thick] (2,0) .. controls (1.25,1) .. (2,2);
    \draw[thick] (0.55,1) -- (1.45,1);
    \node at (3,1) {\Large $=$};
    \node at (4,1) {\LARGE $\frac{1}{\sqrt{2}}$};
    \draw[thick] (5,0) .. controls (6,0.75) .. (7,0);
    \draw[thick] (5,2) .. controls (6,1.25) .. (7,2);
    \node at (8,1) {\Large $-$ \LARGE $\frac{1}{\sqrt{2}}$};
    \draw[thick] (9,0) .. controls (10,0.75) .. (11,0);
    \draw[thick] (9,2) .. controls (10,1.25) .. (11,2);
    \draw[thick,red,dashed] (10,1.45) -- (10, 0.55);
\end{tikzpicture}
\nonumber
\end{align}
\caption{F-symbols of Rep($S_3$)~.}
\label{fig:F_symbols}
\end{figure}

Consider the two-point correlator of the two-cycles encircled by the ${\cal I}_S$ line. We can shrink the line-defect to a point by avoiding both operators, which gives us the quantum dimension of the line $\langle {\cal I}_S \rangle.$ Alternatively, we can make use of the F-symbols \cite{barter2022computing, Bhardwaj:2023idu} in Figure \ref{fig:F_symbols} corresponding to the crossing relations of Rep$(S_3)$ to pull the line-defect through the operators and then shrink it to a point. Pictorially, the two ways of shrinking the defect 
are
\begin{align}
\begin{tikzpicture}
\node at (0,0) {$ \langle \sigma_{[2]} \; \sigma_{[2]} \rangle = \frac{1}{\dim S} \; \langle \; \; \; \sigma_{[2]} \; \;  \sigma_{[2]} \; \; \;  \rangle = \frac{1}{2 \sqrt{2}} \; \langle \; \; \; \sigma_{[2]} \hspace{1.5cm}  \sigma_{[2]} \; \; \;  \rangle  = \frac{\alpha^2}{2 \sqrt{2}} \; \langle \; \sigma^S_{[2]} \hspace{1.5cm}  \sigma^S_{[2]} \; \rangle~, $};
\draw[thick] (-3.225,0) ellipse (0.9 and 0.5);,
\draw[thick] (-0.1,0) circle (0.5cm);
\draw[thick] (1.96,0) circle (0.5cm);
\draw[thick] (0.4,0) -- (1.47,0);
\fill (4.9,0) circle (2pt);
\fill (6.1,0) circle (2pt);
\draw[thick] (4.9,0) -- (6.1,0);
\end{tikzpicture}
\label{sigma2_2point}
\end{align}
Requiring that the correlators on both sides are normalized to $1$, we get $\alpha^2 = 2 \sqrt{2}. $ Similarly, for the two-point correlator of the three-cycle ground states,
\begin{align}
\begin{tikzpicture}
\node at (0,0) {$ \langle \sigma_{[3]} \; \sigma_{[3]} \rangle = \frac{1}{\dim S} \; \langle \; \; \; \sigma_{[3]} \; \;  \sigma_{[3]} \; \; \;  \rangle = \frac{1}{4} \; \langle \; \; \; \sigma_{[3]} \hspace{0.8cm}  \sigma_{[3]} \; \; \;  \rangle  + \frac{1}{4} \; \langle \; \; \; \sigma_{[3]} \hspace{1.5cm}  \sigma_{[3]} \; \; \;  \rangle $};
\draw[thick] (-2.54,0) ellipse (0.9 and 0.5);
\draw[thick] (0.2,0) circle (0.5cm);
\draw[thick] (1.5,0) circle (0.5cm);
\draw[thick] (3.85,0) circle (0.5cm);
\draw[thick] (5.9,0) circle (0.5cm);
\draw[thick, red,dashed] (4.37,0) -- (5.38,0);
\end{tikzpicture} \nonumber \\
\begin{tikzpicture}
\node at (-2.9,0) {$ = \frac{1}{4} \; \langle  \sigma_{[3]} \;   \sigma_{[3]}  \rangle  + \frac{\beta^2}{4} \; \langle \; \sigma^A_{[3]} \hspace{1.5cm}  \sigma^A_{[3]} \; \rangle~, $};
\fill (-0.8,0) circle (2pt);
\fill (-1.96,0) circle (2pt);
\fill (2.1,0) circle (0pt);
\draw[thick, red,dashed] (-0.86,0) -- (-1.9,0);
\end{tikzpicture}
\end{align}
leads to $\beta^2 = 3.$ 

We can now obtain selection rules for the three-point correlators in the presence of the line-defects. The only non-trivial three-point correlators of bare twist operators are $\langle \s_{[2]} \s_{[2]} \s_{[3]} \rangle$ and $\langle \s_{[3]} \s_{[3]} \s_{[3]} \rangle$. To illustrate the procedure we focus on the former. As in the case of two-point functions, the locations of the operators in the three-point correlators are fixed using the $SL(2,\mathbbm{R})$ symmetry. Consider a line-defect ${\cal I}_S$ encircling only $\s_{[3]}.$ This is equivalent to encircling both the $\s_{[2]}$ operators on the sphere as illustrated in Figure \ref{fig:ward223-1}. The second line in the figure is obtained by using the F-symbols in Figure \ref{fig:F_symbols} and the lasso diagram in Figure \ref{fig:lasso}.
\begin{figure}[ht]
\centering
\begin{align}
\begin{tikzpicture}
    \node at (-0.5,0) {$-$};
    \draw (0.5,0.7) -- (0,0) -- (0.5,-0.7);
    \node at (1.5,0.4) {$\s_{[2]} \hspace{0.4cm} \s_{[2]}$};
    \node at (1.5,-0.4){$\s_{[3]}$};
    \draw (2.5,0.7) -- (3,0) -- (2.5,-0.7);
    \node at (3.7,0) {$=$};
    \draw (4.9,0.7) -- (4.4,0) -- (4.9,-0.7);
    \node at (5.9,0.4) {$\s_{[2]} \hspace{0.4cm} \s_{[2]}$};
    \node at (5.9,-0.4){$\s_{[3]}$};
    \draw (6.9,0.7) -- (7.4,0) -- (6.9,-0.7);
    \draw[thick] (5.9,-0.4) circle (0.4cm);
    \node at (8.1,0) {$=$};
    \draw (9.4,0.7) -- (8.9,0) -- (9.4,-0.7);
    \node at (10.4,0.4) {$\s_{[2]} \hspace{0.4cm} \s_{[2]}$};
    \node at (10.4,-0.4){$\s_{[3]}$};
    \draw (11.4,0.7) -- (11.9,0) -- (11.4,-0.7);
    \draw[thick] (10.4,0.4) ellipse (0.9 and 0.5);
\end{tikzpicture} \nonumber \\ \nonumber \\
\begin{tikzpicture}
    \node at (-1,0) {$= $};
    \node at (-0.3,0) {$\frac{1}{\sqrt{2}} $};
    \draw (0.5,0.7) -- (0,0) -- (0.5,-0.7);
    \node at (1.75,0.4) {$\s_{[2]} \hspace{1.1cm} \s_{[2]}$};
    \node at (1.75,-0.4){$\s_{[3]}$};
    \draw (3,0.7) -- (3.5,0) -- (3,-0.7);
    \draw[thick] (0.93,0.4) circle (0.4cm);
    \draw[thick] (2.56,0.4) circle (0.4cm);
    \draw[thick] (1.33,0.4) -- (2.16,0.4);
    \node at (4.2,0) {$= $};
    \node at (4.9,0) {$2$};
    \draw (5.7,0.7) -- (5.2,0) -- (5.7,-0.7);
    \node at (6.95,0.4) {$\s^S_{[2]} \hspace{1.1cm} \s^S_{[2]}$};
    \node at (6.95,-0.4){$\s_{[3]}$};
    \draw (8.2,0.7) -- (8.7,0) -- (8.2,-0.7);
    \draw[thick] (1.33,0.4) -- (2.16,0.4);
    \fill (6.5,0.4) circle (1.5pt);
    \fill (7.4,0.4) circle (1.5pt);
    \draw[thick] (6.5,0.4) -- (7.4,0.4);
    \fill (10,0) circle (0pt);
\end{tikzpicture}\nonumber
\end{align}
\caption{
The Ward identity associated with ${\cal I}_S$ wrapping $\s_{[3]}$. We can shrink the line-defect the other way on the sphere which gives the correlator with both the $\s_{[2]}$ encircled. 
}
\label{fig:ward223-1}
\end{figure}
The correlator $\langle \s_{[2]} \s_{[2]} \s_{[3]} \rangle$ can be expanded using \eqref{eq:g-inv_def_S3} as follows
\begin{align}
    -\langle \s_{[2]} \s_{[2]} \s_{[3]} \rangle &= \frac{-1}{3\sqrt{2}}\langle \; ( \s_{(12)} + \s_{(23)} + \s_{(13)} ) ( \s_{(12)} + \s_{(23)} + \s_{(13)} ) ( \s_{(123)} + \s_{(132)} ) \; \rangle \nonumber \\
    &= -\sqrt{2} \; C_{223}~,
\end{align}
where we have defined $C_{223} \equiv \langle \s_{(12)} \s_{(23)} \s_{(123)} \rangle = \langle \s_{(12)} \s_{(13)} \s_{(132)} \rangle = \langle \s_{(23)} \s_{(13)} \s_{(123)} \rangle.$ Meanwhile, the rightmost correlator in the second line of Figure \ref{fig:ward223-1} can be computed using \eqref{eq:non-gauge_def_S3} as
\begin{align}\label{eq:223_wardcheck}
    2 \langle \s^S_{[2]} \; \begin{tikzpicture}
        \draw[thick] (0,0) -- (0.5,0);
        \fill (0,0) circle (2pt);
        \fill (0.5,0) circle (2pt);
    \end{tikzpicture} \; \s^S_{[2]} \; \s_{[3]} \rangle &= 2 \; \frac{1}{6} \; \frac{1}{\sqrt{2}} \langle (2 \s_{(12)} - \s_{(23)} -\s_{(13)} )  (2 \s_{(12)} - \s_{(23)} -\s_{(13)} ) ( \s_{(123)} + \s_{(132)} ) \rangle \nonumber \\
    &= - \sqrt{2} \; C_{223}  = -\langle \s_{[2]} \s_{[2]} \s_{[3]} \rangle~.
\end{align}
Hence, using the explicit expressions of non-gauge invariant twist fields, we can check the Ward identity of $\mathcal{I}_S$. Similarly, we can also use these defects to calculate the following correlator
\begin{align}
    \begin{tikzpicture}
    \draw (5.2,0.7) -- (4.7,0) -- (5.2,-0.7);
    \node at (6.45,0.4) {$\s^S_{[2]} \hspace{1.4cm} \s^S_{[2]}$};
    \node at (6.45,-0.7){$\s^A_{[3]}$};
    \draw (7.7,0.7) -- (8.2,0) -- (7.7,-0.7);
    \fill (5.85,0.4) circle (1.5pt);
    \fill (7.05,0.4) circle (1.5pt);
    \draw[thick] (5.85,0.4) -- (7.05,0.4);
    \fill (10,0) circle (0pt);
    \draw[thick, red,dashed] (6.45,0.38) -- (6.45,-0.4);
    \fill (6.45, -0.4) circle (1.5pt);
    \node at (9,0) {$=$};
    \node at (11,0) {$-\frac{\sqrt{3}}{2} \langle \s_{[2]} \s_{[2]} \s_{[3]} \rangle~. $};
    \end{tikzpicture}
\end{align}

\begin{figure}[ht]
\centering
\begin{align}
\begin{tikzpicture}
    \node at (-0.5,0) {};
    \draw (0.5,0.7) -- (0,0) -- (0.5,-0.7);
    \node at (1.5,0.4) {$\s_{[2]} \hspace{0.4cm} \s_{[2]}$};
    \node at (1.5,-0.4){$\s_{[3]}$};
    \draw (2.5,0.7) -- (3,0) -- (2.5,-0.7);
    \node at (3.7,0) {$= \frac{1}{2}$};
    \draw (4.9,0.7) -- (4.4,0) -- (4.9,-0.7);
    \node at (5.9,0.4) {$\s_{[2]} \hspace{0.4cm} \s_{[2]}$};
    \node at (5.9,-0.4){$\s_{[3]}$};
    \draw (6.9,0.7) -- (7.4,0) -- (6.9,-0.7);
    \draw[thick] (5.9, 0.15) circle (0.9cm);
    \node at (8.1,0) {$=\frac{\alpha^2}{4}$};
    \draw (9.2,0.7) -- (8.7,0) -- (9.2,-0.7);
    \node at (10.4,0.4) {$\s^S_{[2]} \hspace{1.4 cm} \s^S_{[2]}$};
    \node at (10.4,-0.4){$\s_{[3]}$};
    \draw (11.6,0.7) -- (12.1,0) -- (11.6,-0.7);
    \draw[thick] (10.4,-0.4) circle (0.4cm);
    \draw[thick] (9.75, 0.2) -- (10.11, -0.13);
    \draw[thick] (11.05, 0.2) -- (10.69, -0.13);
    \fill (9.75, 0.2) circle (1.5pt);
    \fill (11.05, 0.2) circle (1.5pt);
\end{tikzpicture} \nonumber \\ \nonumber \\
\begin{tikzpicture}
    \node at (-0.5,0) {$= -\frac{\a^2}{4\sqrt{2}} \; \; $};
    \draw (0.7,0.7) -- (0.2,0) -- (0.7,-0.7);
    \node at (1.85,0.4) {$\s^S_{[2]} \hspace{1.4cm} \s^S_{[2]}$};
    \node at (1.85,-0.4){$\s_{[3]}$};
    \draw (3.1,0.7) -- (3.6,0) -- (3.1,-0.7);
    \draw[thick] (1.25,0.4) -- (2.4,0.4);
    \fill (1.25,0.4) circle (1.5pt);
    \fill (2.4,0.4) circle (1.5pt);
    \node at (4.3,0) {$-\frac{\a^2 \b}{4\sqrt{2}} \; \; $};
    \draw (5.3,0.7) -- (4.8,0) -- (5.3,-0.7);
    \node at (6.45,0.4) {$\s^S_{[2]} \hspace{1.4cm} \s^S_{[2]}$};
    \node at (6.45,-0.7){$\s^A_{[3]}$};
    \draw (7.7,0.7) -- (8.2,0) -- (7.7,-0.7);
    \draw[thick] (1.33,0.4) -- (2.16,0.4);
    \fill (5.85,0.4) circle (1.5pt);
    \fill (7.05,0.4) circle (1.5pt);
    \draw[thick] (5.85,0.4) -- (7.05,0.4);
    \fill (10,0) circle (0pt);
    \draw[thick, red,dashed] (6.45,0.38) -- (6.45,-0.4);
    \fill (6.45, -0.4) circle (1.5pt);
\end{tikzpicture}\nonumber
\end{align}
\caption{
The Ward identity associated with ${\cal I}_S$ encircling all three operators giving rise to a junction. Shrinking the line-defect by avoiding the operator insertions gives us $\langle{\cal I}_S\rangle = 2.$
}
\label{fig:ward223-2}
\end{figure}

\section{Non-universal defects}
\label{sec:non-universal}

In section \ref{sec:universal}, we constructed universal defects labeled by an irreducible representation $R$ of the symmetric group $S_N$. As can be seen from the expression (\ref{td-projector}), the defect operator in each twisted sector is proportional to the identity operator and hence does not depend on the details of the seed CFT. Another way to look at this is that the topological defect one starts with, in the seed theory, is the identity or trivial defect. It is an interesting question whether one can start with a nontrivial defect in the seed theory and construct non-universal defects which depend on the details of the seed theory and the topological defects present in it. Similar questions arise in the construction of D-brane states in symmetric orbifolds, as discussed in  \cite{Gaberdiel:2021kkp,Belin:2021nck}, which we will build upon in the following section.

Here we only construct a simple example of a non-universal defect that is analogous to the ``maximally fractional'' D-brane of \cite{Gaberdiel:2021kkp}. A twisted sector  (\ref{cyclemg}) in the $S_N$ orbifold theory is labeled by a conjugacy class $[g]$ which is fixed by giving the number $l_i$ of cycles of given length $i=1,2,\cdots N$.
We focus on diagonal RCFTs discussed in section \ref{subsec:rcft} as the seed CFT. In particular, we use the defects ${\cal I}_a$  in the RCFT  defined in (\ref{rcft defect}). In the tensor product states, we will use the same boundary state labeled by $a$ for all factors to construct a ``maximally fractional'' defect, which is automatically $S_N$  invariant\footnote{For  discussions on   a more general construction of D-branes in symmetric orbifold CFTs see \cite{Belin:2021nck}.}.  In a twisted sector corresponding to a single $w$ cycle, the doubled boundary state is given by
\begin{align}\label{doubled-b}
\mid B_a\rangle \rangle_{(w)} = \sum_ i {S_{ai}\over S_{0i} }   \sum_{n} \ket{i, n}^{(1)}_{(w)} \otimes \overline{\ket{i,n}}^{(2)}_{(w)}   \sum_{m}   \ket{i,m}^{(2)}_{(w)} \otimes \overline{\ket{i,m}}^{(1)}_{(w)}~,
\end{align}
where the sum over $n,m$ denotes the sum over all the descendants (including fractional Virasoro modes) of the $w$-twisted sector primary $\ket{i}_{(w)}$. Using these boundary states for single cycles, we can construct a defect boundary state which is labeled by the representation $R$ of $S_N$ as well as the choice of  primary $a$ of the RCFT as
\begin{align}\label{doubled-c}
\mid B_a^R\rangle \rangle&= \sum_{[g]\in S_N}  \chi_R([g]) \prod_{j=1}^N \prod_{k_j=1}^{l_j} \mid B_a\rangle \rangle_{(j)}~.
\end{align}
The consistency condition for the topological defects derived by Petkova and Zuber in \cite{Petkova:2000ip} is equivalent to the Cardy condition for the doubled boundary state (\ref{doubled-c}).  The cylinder amplitude for two defect-boundary states is given by
\begin{align}
    Z_{ab}^{R_1 R_2} (q)&=\langle \langle B_a^{R_1}\mid   q^{L^{(1)}_0+\bar L_0^{(1)}-{c\over 12}} \; q^{L^{(2)}_0+\bar L_0^{(2)}-{c\over 12}} \mid B_b^{R_2}\rangle \rangle\nonumber\\
    &=  \sum_{[g]\in S_N} \bar\chi_{R_1}([g]) \; \chi_{R_2}([g])\prod_{j=1}^N \left( \sum_{i}  {S_{ai}^*\over S_{0i} }  {S_{bi}\over S_{0i} }\; \chi_{i}({t/j}) \;\chi_{i}({t/j})\right)^{l_j}~.
 \end{align}
 The Cardy condition states that under modular transformation to the annulus channel, the partition function has a consistent interpretation as an open string partition function, i.e. the primary representations appear with integer multiplicities. Using the modular transformation of the RCFT characters and the Verlinde formula, one obtains
 \begin{align}\label{cardy-a}
    Z_{ab}^{R_1 R_2}(\tilde q) &=   \sum_{[g]\in S_N} \bar\chi_{R_1}([g]) \; \chi_{R_2}([g])\prod_{j=1}^N \Big( \sum_{rs} N_{ab}^{\;\;rs} \; \chi_r (j \tilde t) \; \chi_s (j \tilde t)\Big)^{l_j}~,
 \end{align}
where $N_{ab}^{\;\;rs}$ can be expressed in terms of the fusion coefficients $n^{a}_{\;bc}$ of the seed RCFT
\begin{align}
   N_{ab}^{\;\;rs}&= \sum_{i,k,j} {S_{ai}^* S_{bi} S_{ik}\over S_{0i}}   {S_{kj}^* S_{jr}S_{is} \over S_{0j}}\nonumber \\
   &=\sum_{k} n^k_{ab} \; n_k^{\;rs}~.
\end{align}
Consequently, the $N_{ab}^{\;\;rs}$ are non-negative integers.   The partition function appearing in (\ref{cardy-a})
\begin{align}
   Z_{ab}(\tilde t)= \sum_{rs} N_{ab}^{\;\;rs} \; \chi_r ( \tilde t) \; \chi_s ( \tilde t)~,
\end{align}
contains the characters of the folded RCFT with integer multiplicities and can be written as a trace over a Hilbert space ${\cal H}_{ab}$ in the doubled tensor product of the seed RCFT. Hence, (\ref{cardy-a}) can be written as
\begin{align}
    Z_{ab}^{R_1 R_2}(\tilde q) &=  \sum_{[g]\in S_N} \bar\chi^{R_1}([g]) \;  \chi^{R_2}([g]) \; \tr_{({\cal H}_{ab})^{\otimes N}} \Big( g\;  \tilde q^{L_0-{c\over 24}}\big)~,
\end{align}
and the character sum projects onto the representations of $S_N$ in the $N$-fold tensor product of the folded CFT of $R_1$ and $R_2$. It follows that the open string partition function will be a sum over states with integer multiplicity. The maximal fractional boundary state (\ref{doubled-b}) satisfies the Cardy condition and, therefore, defines a consistent topological defect.

It follows from unfolding the boundary state (\ref{doubled-c}) that the quantum dimension of the maximally fractional defect  ${\cal I}_{a,R}$ labelled by $a,R$ is given by
\begin{align}\label{qdima}
\langle {\cal I}_{a,R}\rangle &= \langle 0 \mid B_a^R\rangle\rangle = \chi_R([1]) \left( {S_{a0}\over S_{00}}\right)^N~,
\end{align}
which is the product of the quantum dimension of the universal defect and the quantum dimensions of the $N$ copies of the topological defect in the RCFT present in the untwisted sector.  Similar considerations as presented above determine the fusion coefficients of the defects as well as their defect Hilbert spaces. 

The construction of the maximally fractional defects can straightforwardly be extended to other seed CFTs including compact bosons, $T^4,$ and $K_3$ sigma models (see e.g. \cite{Frohlich:2006ch,Fuchs:2007tx,Bachas:2008jd,Bachas:2012bj,Brunner:2013ota,Cordova:2023qei,Kostistoappear}), where the RCFT defects in (\ref{doubled-c}) are replaced by topological defects of the seed CFT, which realize more general non-invertible symmetries. We leave the discussion of these defects for future work.

\section{Large-N symmetric orbifold}
\label{sec:large-N}

In this section, we discuss the universal defects of section \ref{sec:universal} in the large-$N$ limit. We provide a criterion for a defect to be non-trivial in this limit. Then, we show that there exist representations that pass this criterion. In the next section, we will argue that in the  AdS$_3$/CFT$_2$ correspondence, these non-trivial defects correspond to topological defects on the worldsheet. 

\begin{figure}[ht]
    \centering
    \begin{tikzpicture}[scale = .7, baseline = 0]
        \coordinate (o) at (0,0);
        \coordinate (d) at (.7,-1);
        \coordinate (m) at (.7,0);
        \coordinate (u) at (.7,1);
        \draw[fill] (o) circle [radius = .05];
        \node[anchor = south] at (o) {$\s_{[g]}$};
        \draw[] (d) -- (u);
        \node[anchor = west] at (d) {$\mathcal{I}_R$};
    \end{tikzpicture}
    \hspace{1.5em}
    $=$
    \hspace{1.5em}
    $\frac{\chi_R([g])}{dim_R}$
    \begin{tikzpicture}[scale = .8, baseline = 0]
        \coordinate (o) at (1,0);
        \coordinate (d) at (-.7,-1);
        \coordinate (m) at (.7,0);
        \coordinate (u) at (-.7,1);
        \draw[fill] (o) circle [radius = .05];
        \node[anchor = south] at (o) {$\s_{[g]}$};
        \draw[] (d) -- (u);
        \node[anchor = west] at (d) {$\mathcal{I}_R$};
    \end{tikzpicture}
    \hspace{1.5em}
    $+ $
    \hspace{1.5em}
    $\sum\limits_{R'\neq \1 } \sum\limits_{i,v}\;  c_{RR'}^i([g]) $
    \begin{tikzpicture}[scale = .7, baseline = 0]
        \coordinate (o) at (1,0);
        \coordinate (d) at (-.7,-1);
        \coordinate (m) at (.7,0);
        \coordinate (u) at (-.7,1);
        \coordinate (e) at (-.7,0);
        \draw[fill] (e) circle [radius = .05];
        \node[anchor = south west] at (e) {$v$};
        \draw[fill] (o) circle [radius = .05];
        \node[anchor = south] at (o) {$\s_{[g]}^{R',i}$};
        \draw[] (d) -- (u);
        \draw[] (e) -- (o);
        \node[anchor = west] at (d) {$\mathcal{I}_R$};
        \node[anchor = south] at (0.3,-0.7) {$\mathcal{I}_{R'}$};
    \end{tikzpicture}
    \caption{Action of the topological defect $\mathcal{I}_R$ on twist operators. The vector $v$ is an element in Hom$(R\otimes R, R')$. The sum on the right-hand side is over $R'$ with $N_{RR}\;^{R'} \neq 0$ . We have omitted the orientation of the defect since all the representations of $S_N$ are real. }
    \label{fig:Action_on_s}
\end{figure}

Consider the topological defect $\mathcal{I}_R$, the general action on twist operators is shown in Figure \ref{fig:Action_on_s}. This depends on the Kronecker coefficients $N_{RR}\;^{R'}$ as well the constants $c_{RR'}^i([g])$. The former determines the representations $R'$ that appear in the sum. In principle, to determine the constants $c_{RR'}^i([g])$ we need to know both the  Kronecker coefficients as well as the F-symbols. At present, it is an open problem in mathematics to find explicit expressions for the Kronecker coefficients for arbitrary representations. In fact, it is known that the Kronecker coefficients in general are unbounded for large-$N$ \cite{pak2015complexity}.

For a defect to be non-trivial in the large-$N$ limit, it needs to satisfy\footnote{Note the characters satisfy $\chi_R([g]) \leq dim_R $ for any representation.}
\begin{equation}
     \lim_{N \rightarrow \infty}\frac{\chi_R([g])}{dim_R}  < 1~.
    \label{eq:constraint}
\end{equation}
If this is not the case, the defect commutes with all the operators, and the contributions of non-local operators on the right-hand side of Figure \ref{fig:Action_on_s} are subleading \cite{Chang:2018iay}.   

Let us now comment on what representations lead to non-trivial Ward identities in the large-$N$ limit. Recall that as is the case for the conjugacy classes, representations of $S_N$ are also labeled by partitions of $N$ \footnote{Here we are using a different notation to label partitions of $N$ compared to section \ref{sec:universal} where $l_i$ is the number of $r_i$'s that satisfy $r_i = i$.}

\begin{equation}
    R: \quad (r_1, r_2, \dots), \quad r_1+r_2+\dots = N~.
\end{equation}
We can associate a Young tableau to each representation where the number of boxes in the row $i$ is equal to $r_i$. For a representation $R,$\footnote{The representation needs to have a well-defined limit, see for example \cite{Borodin}.} the character of a conjugacy class with a single cycle has the following limit \cite{okounkov1998representations, Borodin}

\begin{align}
    \lim_{N \rightarrow \infty} \frac{\chi_R([n])}{dim_R} = \sum_i \a_i ^n - (-1)^n\sum_i \b_i ^n~,
\end{align}
where $\a_i =  \lim\limits_{N \rightarrow \infty} \frac{r_i}{N}$ is the number of boxes in row $i$ divided by $N$. Similarly, $\b_i$ is the number of boxes in the column $i$ divided by $N$. 
As an example, let us look at the standard representation of $S_N$, which corresponds to the Young tableau
\[
\begin{ytableau}
    ~ &  &  & \cdots & \\
     ~& \none & \none & \none & \none      
\end{ytableau}
\]
with the corresponding partition being $(N-1,1)$. We see that

\begin{equation}
    \a = \lim_{N \rightarrow \infty}\left(\frac{N-1}{N}, \frac{1}{N}, 0, 0, \dots\right) = (1,0,0,\dots)~,
\end{equation}
and
\begin{equation}
    \b = \lim_{N \rightarrow \infty}\left(\frac{2}{N}, \frac{1}{N}, \frac{1}{N}\dots\right) = (0,0,0,\dots)~.
\end{equation}
Hence, in the large-$N$ limit $\frac{\chi_S([g])}{dim_S}\rightarrow 1$. This can also be derived by calculating the character of the standard representation  for finite $N$ using the Frobenius formula. Therefore, we conclude that the Ward identities of the standard representations are trivial in the large-$N$ limit and only constrain subleading contributions coming from higher genus covering spaces.

Examples of representations with nontrivial ratio $\frac{\chi_R([g])}{dim_R}$ in the large-$N$-limit are Young diagrams with two rows and diagrams of hook-shape. For both of the above, the Kronecker coefficients $N_{RR}\;^{R'}$ are known \cite{REMMEL1989100,10.36045/bbms/1103408635, rosas2000kronecker}.  For example, for the Young diagram with two rows

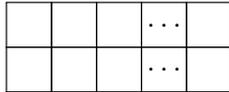
\[
\begin{ytableau}
    ~ &  &  & \cdots & \\
    ~&  &  & \cdots & \\    
\end{ytableau}
\]
we have
\begin{equation}
    \a =  \left(\frac12,\frac12,0,\dots\right), \quad  \b = (0,0,0,\dots)~,
\end{equation}
and therefore 
\begin{equation}
    \frac{\chi_R([n])}{dim_R}\rightarrow \frac{1}{2^{n-1}}~.
\end{equation}
Thus, we expect the defect corresponding to this representation to be non-trivial in the large-$N$ limit. 

\section{Non-invertible symmetries  and AdS$_3$/CFT$_2$ duality}
\label{sec:non-inv-ads-cft}
 The tensionless limit of the type II $AdS_3\times S^3\times \mathcal M_4$ background with NS-NS flux is a holographic setup where the non-invertible symmetries of the $S_N$  orbifold CFT are realized. In this section, we take some first steps in constructing the holographic duals of these defects \footnote{See the recent paper \cite{Heckman:2024obe} for related discussions.}.

\subsection{Tensionless Limit}

In recent years, an example of an exact AdS/CFT duality  has been established, relating type II string theory on $AdS_3\times S^3\times \mathcal M_4$ with $k=1$ units of NS-NS flux to the symmetric orbifold CFT $\mathcal M_4^{\otimes N}/S_N$ in the large-$N$ 
limit. Here, $\mathcal M_4$ is the CFT defined by the supersymmetric sigma-model with target space  $T^4$ or $K_3$. The characteristic feature of this duality is that the string theory is 
defined on highly curved spaces far away from the supergravity limit, yet a tractable worldsheet CFT exists. The worldsheet description is based on a product of a 
$PSU(1,1|2)$ WZW model times a twisted $\mathcal M_4$ sigma model \cite{Gaberdiel:2018rqv,Berkovits:1999im}. The agreement of the 
spectra on the AdS and CFT side was demonstrated in \cite{Eberhardt:2018ouy,Gaberdiel:2018rqv}, correlation functions were identified in 
\cite{Dei:2019iym,Dei:2020zui}, the torus partition function was related to thermal AdS in \cite{Eberhardt:2020bgq}, and D-branes on both sides were related in \cite{Gaberdiel:2021kkp}.  The identification of states on both sides relates the vertex operators ${\cal O}_h^{(w)}(x)$ for a CFT state in a single cycle twisted sector labeled by $w$ and a seed CFT state labeled by $h$, to a worldsheet vertex operator $V_h^w(x;z)$ where $w$ now denotes the $w$-spectral flow sector. This duality relates string amplitudes to the orbifold worldsheet correlation function
\begin{align}\label{dictionary}
    &\big\langle {\cal O}_{h_1}^{(w_1)}(x_1)\; {\cal O}_{h_2}^{(w_2)}(x_2)\cdots {\cal O}_{h_n}^{(w_n)}(x_n)\big\rangle_{S^2} \nonumber\\
    =&\;\int_{{\cal M}_{g,n}} d\mu\big\langle V_{h_1}^{w_1}(x_1;z_1)V_{h_2}^{w_2}(x_2;z_2)\cdots V_{h_n}^{w_n}(x_n;z_n)\big\rangle_{\Sigma_{g,n}}~.
\end{align}
The equality of both sides hinges on a remarkable localization property of the string path integral. The integral over the moduli space ${\cal M}_{g,n}$ only contributes if a holomorphic covering map exists from the worldsheet $\Sigma_{g,n}$ to the spacetime $S^2$. In this map, the vertex operator locations $z_i$ on the worldsheet are mapped to the locations $x_i$ of the CFT operators. These points have ramification indices $w_i$ for $i=1,2,\cdots,n$.   The covering map behaves like
\begin{align}\label{covering-map}
    \Gamma(z) = x_i+ a_i^\Gamma(z-z_i)^w_i+ {\cal O}\big( (z-z_i)^{w_i+1}\big)~,
\end{align}
near $z=z_i$ and hence close to $z_i$ describes a $w_i$-fold cover. The fact that the string worldsheet path integral localizes was shown using worldsheet Ward identities in \cite{Eberhardt:2019ywk} (see \cite{Knighton:2023mhq,McStay:2023thk,Hikida:2023jyc,Lerche:2023wkj,Knighton:2024qxd} for further discussions). In the following, we will focus on correlation functions at leading order in large-N, where both the string worldsheet and the target space are two spheres.  This map allows us to reverse engineer the topological defect on the string worldsheet.

We recall that the topological defect operators associated with Rep$(S_N)$ given in (\ref{td-projector}) are projectors on the twisted sectors of the $S_N$ orbifold, weighted by the character of the representation
 \begin{align}\label{td-projector-b1}
{\cal I}^{\rm CFT} _R&=  \sum_{[g]}  \chi_R([g]) P_{[g]}\bar { P}_{[g]} ~.
\end{align}

Single string states created by the vertex operators $V_h^w(x,z)$ in the spectral flow sector $w$ correspond to single cycle twisted states of length $w$.  Limiting ourselves to the single cycle twisted sector, a natural proposal for a worldsheet is the projector
\begin{align}\label{td-projector-b}
{\cal I}^{\rm ws} _R&=  \sum_{w}  \chi_R([w]) { P}_{w}\bar { P}_{w} ~,
\end{align} 
where ${ P}_w$ is a projector on the $w$-spectral flow sector and $\chi_R([w])$ is the $S_N$ character of a single cycle conjugacy class of length $w$ in representation $R$.

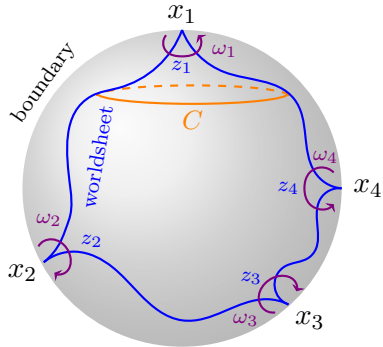
\begin{figure}[ht]
 \centering
\begin{tikzpicture}[scale = .7, baseline = 0]
  \shade[ball color = gray!40, opacity = 0.4] (0,0) circle (3cm);
  \draw[orange,thick] (-1.6, 1.8) arc (180:360: 1.8 and 0.2);
  \draw[dashed,thick,orange] (2, 1.75) arc (0:180: 1.8 and 0.2);
\draw [thick,blue] (0,3) to [ curve through ={(-1.2, 1.9) . . 
(-1.7, 1.75)..
..(-2.1, 0)..(-2.1, -0.5)
..(-2.18, -0.75)..(-2.28, -1)}] (-2.6, -1.4);
 \draw [thick,blue] (0, 3) to [ curve through ={(0.8, 2.2)..(2.1, 1.7)..(2.5, 0.3)..(2.7, 0.1)}] (3, 0);
 \draw [thick,blue]  (3, 0) to [ curve through ={(2.6,-0.2).. (2.4,-1.3)..(1.9,-1.5)}] (2, -2.2);
 \draw [thick,blue]  (-2.6, -1.4) to [ curve through ={(-2, -1.2)..(-1.3,-1.5)..(-0.8,-2)..(0,-2.5)..(1.6, -2.1)}] (2, -2.2); 
   \draw[-{Stealth[scale length=0.3]}, violet, thick] (-0.3, 2.9)  to 
   [ curve through ={ (-0.3,2.6)..(0,2.5)..(0.3,2.6)}](0.3, 2.9) ;
    \draw[-{Stealth[scale length=0.3]}, violet, thick] (-2.7, -1.1)  to [ curve through ={(-2.6, -1)..(-2.1,-1.3)..(-2.2,-1.6)}] (-2.4, -1.6);
    \draw[-{Stealth[scale length=0.3]}, violet, thick](1.75, -2.35)  to [ curve through ={(1.5, -2.25)..(1.7,-1.7)..(2.16,-1.85)}](2.15, -1.95) ;
    \draw[-{Stealth[scale length=0.3]}, violet, thick] (2.85, 0.3)  to 
      [ curve through ={ (2.4, 0.3)..(2.3,0)..(2.4,-0.3) }]
      (2.85, -0.3) ;
    \node at (0,3.3) {$x_1$};
     \node at (3.5 ,0) {$x_4$};
     \node at (2.4, -2.5) {$x_3$};
     \node at (-3, -1.6) {$x_2$};
    \node[font=\small] at (0.2, 1.3) {$\color{orange}{C}$};
    \node[font=\scriptsize] at (0.8, 2.6) {$\color{violet}{\omega_1}$};
     \node[font=\scriptsize] at (-2.5, -0.6) {$\color{violet}{\omega_2}$};
      \node[font=\scriptsize] at (1.2, -2.5) {$\color{violet}{\omega_3}$};
       \node[font=\scriptsize] at (2.7, 0.6) {$\color{violet}{\omega_4}$};
    \node[font=\scriptsize] at (0, 2.3) {$\color{blue}{z_1}$};
     \node[font=\scriptsize] at (-1.7, -1) {$\color{blue}{z_2}$};
      \node[font=\scriptsize] at (1.3, -1.7) {$\color{blue}{z_3}$};
       \node[font=\scriptsize] at (2, 0) {$\color{blue}{z_4}$}; 
      \path (-1.6, 0.6) node[rotate=80, blue] {\scriptsize{worldsheet}};
       \path (-2.6, 2) node[rotate=55] {\scriptsize{boundary}};
\end{tikzpicture}
\caption{Contour $C$ of operator $I$ enclosing a vertex operator localized at $x_1$.}
\label{fig:localization}
\end{figure}

Such a projector\footnote{Another proposal for $I[C]$ is the boundary central term ${\cal I}=\oint {da\over 2\pi i} {\partial_z\gamma\over\gamma}$ \cite{Giveon:1998ns,Kutasov:1999xu,Giveon:2001up,Kim:2015gak,Eberhardt:2019qcl}. This operator gives the winding number of the covering map $\Gamma(z)$, but the winding number is only defined with respect to a base point ($z=0$ in the given formula) and  does not lead  to a viable worldsheet version of the topological defect.} can be constructed as an operator on the worldsheet using the 
 free field 
Wakimoto construction \cite{Wakimoto:1986gf} of the $SL(2,\mathbbm{R})$  part of the worldsheet sigma model.  As an initial step, we define $I[C]$ as a contour integral over the worldsheet bosonic field $\Phi(z)$\footnote{It has been shown in appendix C of \cite{Eberhardt:2019ywk} that the commutator of $\partial_z \Phi$  with the spacetime Virasoro generators is a total derivative, hence $I(C)$ is topological from the spacetime CFT perspective.}
\begin{align}\label{I-def1}
  I[C]= -2 \oint_C {dz \over 2\pi i } \;\partial_z \Phi~,
\end{align}
where the contour $C$ is always oriented counterclockwise.
The localization of the worldsheet path integral \cite{Eberhardt:2019ywk,McStay:2023thk,Knighton:2023mhq}, pictorially shown in Figure \ref{fig:localization}, leads to a very special property of the  $\partial_z\Phi(z)$ field when inserted in string correlation functions \cite{Eberhardt:2019ywk}
\begin{align}\label{local1}
    &\langle \big(-2 \partial\Phi(z) \big)V_{h_1}^{w_1}(x_1;z_1)\cdots V_{h_n}^{w_n}(x_n;z_n)\big\rangle
    = \frac{\partial_z^2 \Gamma(z)}{\partial_z\Gamma(z)} \langle V_{h_1}^{w_1}(x_1;z_1)\cdots V_{h_n}^{w_n}(x_n;z_n)\big\rangle~.
\end{align}
The expectation value of $I(C)$ can then be evaluated using the following identity for the covering map $\Gamma$ 
\begin{align}\label{local2}
   \frac{\partial_z^2 \Gamma(z)}{\partial_z\Gamma(z)} = \sum_{i=1}^n \frac{w_i-1}{z-z_i}-\sum_{a=1}^{N-1} \frac{2}{z-z^*_a}~.
\end{align}
Here, $z_a$ are the location of the ``secret poles,'' which in general depend on the location of the CFT operators $x_i$ and are necessary for the existence of the covering map. We have placed the $N$-th secret pole at $z=\infty,$ which is always possible and doing so cancels the background charge of $\Phi$ at infinity as discussed in \cite{Eberhardt:2019ywk}. It is easy to verify that the residue of $\partial^2_z\Gamma/\partial_z \Gamma$ at infinity is zero using the Riemann-Hurwitz formula. The value of $I[C]$  for a given contour can now be determined using the residue theorem.

We can now give a proposal for  ${\cal I}^{\rm ws}_R[C]$, i.e. the worldsheet version of the topological defect for contour $C$
\begin{align}\label{ws-proj1}
 {\cal I}^{\rm ws}_R[C] &=\sum_{w}\chi_R(w) \delta_{w- I[C]} \nonumber\\
&= \sum_{w=1}^\infty \chi_R(w) \int_0^{2\pi} {d\alpha\over 2\pi} \exp\Big(  i \alpha\big(w -1 - I[C]\big)\Big)~.
\end{align}

Thanks to the localization formula \eqref{local1} and \eqref{local2},
it is easy to see that (\ref{I-def1}) evaluates to $w_i-1$  for a countour which encircles only $z_i$ and is independent of small deformations of the contour.  Hence, (\ref{ws-proj1}) is acting like the projector (\ref{td-projector-b}) when encircling a single vertex operator.
In particular, if the contour
  is inserted around a vertex operator in the untwisted sector  (this can be the ground state itself), we have $w_i=1$  and $I[C]$ evaluates to zero and hence  ${\cal I}^{\rm ws}_R[C]$ evaluates to $\chi^R([1])$, i.e. the quantum dimension of the operator.

Consider now a two-point function of twist operators on the boundary. The dual correlator is a two-point function of spectrally flowed vertex operators. It is not too difficult to check that the value of ${\cal I}^{\rm ws}_R[C]$ around any of the two insertions is the same. This is consistent with inserting the corresponding operator ${\cal I}_R^{CFT}$ on the boundary around either operators which gives the same value. The more interesting case is when we insert the operator ${\cal I}_R^{CFT}$ around a line that does not enclose any operator. This gives the dimension of the representation. On the worldsheet, the corresponding operator ${\cal I}^{\rm ws}_R[C]$ will be along a curve with no vertex operator inside giving again the dimension of the representation. However, in this case, the contour integral in $I[C]$ with no poles inside is not equal to the sum of the contour integrals around each of the vertex operators.
This is because of the the presence of the secret poles in \eqref{local2}.
On the worldsheet, we interpret this as the result of the lasso effect described in Figure \ref{fig:non.inv.lines}.

The sign (alternating) defect is an example of an invertible defect for which $\chi^A (w) =(-1)^{w+1}$. When inserted on the boundary, the defect enforces the selection rule that $\sum_i {w_i}$ is odd. Since the character of the sign representation on single cycles of length $w$ satisfies $\chi_A(w+2) = \chi_A(w)$, we see that ${\cal I}^{\rm ws}_R[C]$ is blind to the secret poles (they contribute $-2$ to the contour integral) and, therefore, the contour integral can be written as a sum of contour integrals around the ramification points reproducing the correct Ward identity on the worldsheet. 
The same line of reasoning can be applied to the case of three-point functions reproducing the corresponding Ward identities on the worldsheet.

\subsection{Deformation away from  the $k=1$ and the  supergravity limit}

Turning on RR-fluxes in the bulk corresponds to deforming the symmetric orbifold CFT by an exactly marginal deformation \cite{Burrington_2013, Gaberdiel:2015uca,Guo_2020,Keller_2020,  Eberhardt:2021vsx,Balthazar:2021xeh,Apolo_2022, Fiset:2022erp, Eberhardt_2022, Gaberdiel:2023lco,Knighton:2024qxd}. This deformation is proportional to a two-cycle twist field which we denote by $\s_{[2]}$. In the following, we argue that this deformation breaks all the non-invertible universal defects associated with the representations of $S_N$. 

Consider the action of a topological defect $\mathcal{I}_R$ on the twisted state created by $\s_{[2]}$
\begin{align}
    \mathcal{I}_R \ket{\s_{[2]}} &= \chi_R([2]) \ket{\s_{[2]}}~.
\end{align}

The deformation breaks all the defects if and only if $\chi_R([2]) \neq dim_R$. Below we prove that no defect satisfies this and, therefore, all the universal defects are broken in the supergravity limit.  

Consider an element $g$ in some representation $R$. Since all finite groups are unitariazable, this is a $dim_R\times dim_R$ matrix whose eigenvalues satisfy $\abs{\l_i} =1$. Hence, 
\begin{align}
    \chi_R(g) &= \sum_{i=1}^{dim_R} \l_i~, \quad \abs{\l_i}=1~.
\end{align}
We see that $\chi_R(g) = dim_R$ if and only if  $\l_i = 1$ $\forall$ $i$ which is true only for the trivial representation. Hence, we conclude that any marginal deformation involving twist operators breaks all the universal defects. This means that all nontrivial symmetries do not survive the deformation away from the $k=1$  point. 

\section{Discussion}
\label{sec:Discussion}

In this paper, we investigated certain aspects of the universal topological defects of the symmetric orbifold CFT $\mathcal M^{\otimes N}/S_N$. These defects are universal in the sense that they do not depend on the details of the seed CFT, and exist in any $S_N$ orbifold theory. Using modular transformations of torus partition function, we determined universal defect operators on which the topological lines can end. These operators are built on the twisted sector ground states. These states are derived from the vacuum of the seed CFT and hence do not depend on the details of the seed CFT other than the central charge and the conjugacy class labeling the twisted sector.
 
The non-invertibility of the symmetry leads to Ward identities which relate correlation functions of local operators to (sums of) correlation functions of defect and local operators with lines as well as junctions inserted. We illustrated this in the simplest example which exhibits non-ivertible symmetries, namely the $S_3$ symmetric orbifold.

In addition, we constructed an example of a non-universal defect which is built from a topological defect in the seed CFT (which we took to be diagonal RCFTs with Verlinde lines). This ``maximally'' fractional defect satisfies the Cardy constraints and it would be interesting to see whether more general solutions (maybe along the lines of similar constructions for D-branes in symmetric orbifolds in \cite{Belin:2021nck}) are possible. 
The maximally fractional defect we constructed in the present paper can be, in some sense, viewed as a product of the RCFT and symmetric orbifold defects.

We provided a criterion that determined which universal defects are non-trivial in the large-$N$ limit. This was possible due to the known asymptotic behavior of $S_N$ characters. It would be particularly interesting to study the Ward identities of non-trivial defects in the large-$N$ limit. At this moment, this task looks daunting since there are no explicit expressions for both the F-symbols as well as the Kronecker coefficients of $S_N$ to our knowledge. 

The universal defects of the orbifold CFT are interesting in the context of AdS$_3$/CFT$_2$ correspondence. We argued that these defects correspond to defects on the worldsheet, a case distinct from previous examples where boundary topological defects correspond to $D$-branes in the bulk. This is a peculiarity of the tensionless limit. This idea has appeared recently in other works \cite{Heckman:2024obe,Kaidi:2024wio}. On the boundary, passing an operator though a topological defect generally produces non-local operators. It would be interesting to understand the holographic dual of these operators. 

Away from the orbifold point, we showed that the universal defects are broken explicitly by a marginal deformation. Hence, these defects do not generally exist in the supergravity limit. It would be interesting to understand what happens when the deformation parameter is infinitesimal and what role these softly broken defects might play in holography.

In this work, we focused on the tensionless limit with $k=1$. This corresponds to the smallest possible radius for AdS$_3$. For higher values of $k,$ it is believed that the boundary theory is tensored with a Liouville factor but remains an orbifold CFT \cite{Eberhardt:2019qcl}. Hence, we expect our universal defects to be present for higher values of $k$ as well. It would interesting to understand the holographic dual of the universal defects in the large $k$ limit where semi-classical supergravity computations are more reliable. 

The entanglement entropy in the presence of topological defects is an interesting observable that has been calculated in some examples for topological defects \cite{Brehm:2015plf,Gutperle:2015kmw}. We plan to generalize these calculations to the topological defects constructed in the present paper \cite{ustoappear} and it would be interesting to understand their implications for holography.

\acknowledgments
The work of M.G. was supported, in part, by the National Science Foundation under grant PHY-2209700.  M.G is grateful to the Department of Physics and Astronomy, Johns Hopkins University for hospitality while this paper was finalized.
M.G., Y.L and D.R. are grateful to the Bhaumik Institute for support. K.R. is supported by the Simons Collaboration on Global Categorical Symmetries and also by the NSF grant PHY-2112699.

\appendix
\section{Defect Hilbert spaces in $S^3$ orbifold}
\label{appendix:a}

In this appendix, we verify the counting of defect operators in section \ref{sec:s3example}. From \eqref{eq:Grand_can}, the torus partition function of the $S_3$ symmetric orbifold  is
\begin{align}
    Z(\tau) &=\frac{1}{6}Z(\tau)^3 +  \frac{1}{3} Z(3\tau) +  \frac{1}{2} Z(\tau)Z(2\tau) + \frac{1}{2} Z(\tau)\left( Z\left(\frac{\tau}{2} \right) + Z\left(\frac{\tau + 1}{2}\right)\right) \nonumber\\
&+ \frac{1}{3}\left(  Z\left(\frac{\tau}{3} \right) + Z\left(\frac{\tau + 1}{3}\right) + Z\left(\frac{\tau + 2}{3}\right)\right),
\end{align} 
where for simplicity we only denoted the dependence on $\tau$, and the dependence on its complex conjugate ${\bar \tau}$  is implied. The partition function of the seed CFT can be divided into the vacuum Virasoro module and the excited states
\begin{align}
    Z(\tau) = \frac{q^{-\frac{c}{24}} {\bar q}^{-\frac{c}{24}}}{\prod \limits_{n=2}^\infty(1-q^n)\prod\limits_{n=2}^\infty(1-\bar q^n)} + \text{excited states}~,
    \label{eq:seedZ}
\end{align}
where $q(\tau)= e^{2\pi i \tau}$.
The vacuum states lead to the universal sector of the symmetric orbifold CFT created by the bare twist operators $\s_{[g]}$.

Let us denote the scaling dimension of primary operators in the seed CFT by $h_{seed}$.  From the partition function, it can be seen that the bare twist operators for a single cycle with length $k$ have scaling dimensions,
\begin{align}
 h^{(bt)}_k = \frac{c}{24}(k-\frac{1}{k})~,
\end{align}
where `bt' is the abbreviation for bare twists. In the $S^3$ orbifold, $k=1, 2, 3$. 
More generally, the scaling dimension in the single $k$-cycle twisted sector is 
\begin{align}
 h = \frac{c}{24}(k-\frac{1}{k}) + \frac{h_{seed}}{k}~,
\end{align}

Let us now consider the defect Hilbert spaces $\mathcal H_{A}, \mathcal H_{S}$ assosciated to the defects of section \ref{sec:s3example}. Using (\ref{twsitedpf}) and (\ref{characterS3}), we can write down the torus partition function with a defect $\mathcal I_A$ inserted along the spacial cycle,
\begin{align}
    Z_A(\tau) =&\frac{1}{6}Z(\tau)^3 +  \frac{1}{3} Z(3\tau) +  \frac{1}{2} Z(\tau)Z(2\tau)-\frac{1}{2} Z(\tau)\left( Z(\frac{\tau}{2} ) + Z(\frac{\tau + 1}{2})\right) \nonumber\\
&+ \frac{1}{3}\left(  Z(\frac{\tau}{3} ) + Z(\frac{\tau + 1}{3}) + Z(\frac{\tau + 2}{3})\right)~.
\end{align}
Under modular transformation (\ref{modular-a1}), the defect partition function becomes 
\begin{align}
    Z^A(\tau') =&  \frac{1}{6}Z(\tau)^3 -\frac{1}{2}Z(\tau) Z(2\tau)  + \frac{1}{3}  Z(3\tau) + \frac{1}{2}Z\left(\tau\right)\left( Z\left(\frac{\tau}{2}\right)  - Z\left(\frac{\tau + 1}{2}\right) \right)\nonumber\\
     &  + \frac{1}{3}\left(  Z\left(\frac{\tau}{3} \right) + Z\left(\frac{\tau + 1}{3}\right) + Z\left(\frac{\tau + 2}{3}\right)\right)~.
     \label{ZA}
\end{align}
Using \eqref{eq:seedZ} we can expand the above in powers of $q$ and $\bar q$.
For the 2-cycle twisted sector, 
\begin{align}
   & \frac{1}{2} Z(\tau)\left( Z\left(\frac{\tau}{2}\right) - Z\left(\frac{\tau + 1}{2}\right)\right) =  
   (q\bar q)^{-\frac{c}{16}}(q^{\frac{3}{2}} + \bar q^{\frac{3}{2}}+  q^\frac{3}{2} \bar q +   q  \bar q^\frac{3}{2} + 3 q^2 \bar q^{3/2} + 3 q^{3/2} \bar q^2 \dots)~,
\end{align}
and hence there is no bare twist-operator in $\mathcal{H}_A$ verifying \eqref{eq:non-gauge_def_S3}. Similarly for the 3-cycle we have that

\begin{align}
   & \frac{1}{3}\left(  Z\left(\frac{\tau}{3} \right) + Z\left(\frac{\tau + 1}{3}\right) + Z\left(\frac{\tau + 2}{3}\right)\right) 
    =(q\bar q)^{-\frac{c}{72}}
 (1+ q  +  q^{2/3}\bar q^{2/3} + \bar q + q \bar q + \dots)~.
\end{align}
Finally for the 1-cycle sector,
\begin{align}
    \frac{1}{6}Z(\tau)^3 +  \frac{1}{3} Z(3\tau) -  \frac{1}{2} Z(\tau)Z(2\tau)
    = (q\bar q)^{-\frac{c}{8}}(q^2\bar q^2 +q^3\bar q^2 + q^2\bar q^3 +   q^3 \bar q^3+3 q^2 \bar q^4+ 3 q^4 \bar q^2 \dots)~.
\end{align}
which means that the is no operator with zero scaling dimensions and therefore $\text{Hom}(\mathcal{I}_A, \mathcal{I}_A)$ is trivial.

Similarly, we can study the ${\cal I}_S$ defect Hilbert space. There is no identity operator in the  $\mathcal I_S$ defect Hilbert space. There is a bare 2-cycle and no bare 3-cycle twist operators, in agreement with \eqref{eq:non-gauge_def_S3}. 

\newpage

\bibliographystyle{JHEP}
\bibliography{bibliography}
\end{document}